%% file: FlowILP.tex
\documentclass[10pt]{article}
\usepackage{etex}
\usepackage[utf8]{inputenc}
\usepackage{cite}
\usepackage{graphicx}
\usepackage{amsmath}
\usepackage{amssymb}
\usepackage{changepage}
\usepackage[noend,noline,ruled]{algorithm2e}
\DontPrintSemicolon
\SetAlCapHSkip{0em}
\usepackage{subcaption}
\usepackage[numbers, sort&compress]{natbib}
\usepackage{hyperref}
\usepackage{siunitx}
\usepackage[dvipsnames, svgnames, table]{xcolor}
\usepackage{booktabs}
\usepackage{mathtools}
\usepackage{multicol}
\usepackage{amsthm}
\usepackage{longtable}
\usepackage{colortbl}
\usepackage{graphbox} 
\theoremstyle{plain}
\newtheorem*{proposition}{Proposition}
\theoremstyle{definition}
\newtheorem*{definition}{Definition}
\usepackage{grffile}

\newcommand{\mset}[1]{\left\{\!\!\left\{#1\right\}\!\!\right\}}
\newcommand{\ch}[1]{\ensuremath{\mathrm{#1}}}
\newcommand{\carrow}{\mbox{\boldmath{\ensuremath{\,\longrightarrow\,}}}}
\newcommand{\carrowBoth}{\mbox{\boldmath{\ensuremath{\,\longleftrightarrow\,}}}}

\usepackage{tikz}
\usepackage{chemfig}

\usepackage{mod}
\tikzset{eMorphism/.style={->, >=stealth}}
\tikzset{eIsomorphism/.style={double, double equal sign distance}}

\tikzset{vDot/.style={fill,minimum size=0, inner sep=1pt, circle, outer sep=0}}
\tikzset{eDot/.style={}}
\tikzset{eDotMorphism/.style={->, >=stealth, dashed, red}}
\tikzset{gDot/.style={node distance=1em and 1em}}
\tikzset{gCat/.style={node distance=2em and 2em}}
\tikzset{vCat/.style={draw, minimum size=2.5em}}
\newcommand\dpoRuleRaw[3]{%
    \begin{tikzpicture}[gCat, baseline={([yshift=-0.5ex]K.center)}]
        \node[vCat, label=above:$L$] (L) {#1};
        \node[vCat, label=above:$K$, right=of L] (K) {#2};
        \node[vCat, label=above:$R$, right=of K] (R) {#3};
        \draw[eMorphism] (K) to (L);
        \draw[eMorphism] (K) to (R);
    \end{tikzpicture}%
}
\newcommand\dpoRule[4][]{%
    \dpoRuleRaw{\includegraphics[#1]{#2}}{\includegraphics[#1]{#3}}{\includegraphics[#1]{#4}}
}

\usetikzlibrary{shapes,positioning,arrows,calc,matrix,shadows,decorations.markings}
\tikzset{hnodeNoDraw/.style={ellipse,minimum height=12, minimum width=17}}
\tikzset{hnode/.style={draw,hnodeNoDraw}}
\tikzset{tnode/.style={draw,circle,fill=black,inner sep=0,minimum size=3,text height=0ex,text depth=0ex}}
\tikzset{hedgeNoDraw/.style={rectangle,minimum height=8, minimum width=8}}
\tikzset{hedge/.style={draw,hedgeNoDraw}}
\tikzset{edge/.style={->,>=stealth',thick}}
\tikzset{tedge/.style={->,>=stealth'}}
\newcommand{\multiedge}[4][20]{
\pgfmathsetmacro{\num}{#4-1}
\foreach \i in {0, ..., \num}{
	\pgfmathsetmacro{\angleSep}{#1}
	\pgfmathsetmacro{\angle}{\angleSep*(\i-#4/2+.5)}
	\draw[edge] (#2) to [bend right=\angle] (#3);
}
}

\newcommand{\removeStuff}[1]{}
\newcounter{incFigCounter}
\stepcounter{incFigCounter}
\newcommand{\incFig}[1]{%
\includegraphics{incFig/\arabic{incFigCounter}.pdf}%
\stepcounter{incFigCounter}%
}

\newcommand\computerDesc{%
Intel\textregistered{} Core\textsuperscript{TM} i7-6700 CPU (\SI{3.40}{\giga\hertz})%
}

\title{%
Chemical Transformation Motifs\\
--- Modelling Pathways as Integer Hyperflows
}

\usepackage{authblk}
\usepackage{marvosym}
\newcommand\corr{\text{\Letter}}
\newcommand\email[1]{\texttt{#1}}

\author[1-4]{Jakob L.\ Andersen}
\author[4,10]{Christoph Flamm}
\author[2,\corr]{Daniel Merkle}
\author[4-9]{Peter F.\ Stadler}
\affil[1]{Earth-Life Science Institute, Tokyo Institute of Technology, Tokyo 152-8550, Japan}
\affil[2]{Department of Mathematics and Computer Science, University of Southern Denmark, Odense M DK-5230, Denmark
    \email{daniel@imada.sdu.dk}}
\affil[3]{Research group Bioinformatics and Computational Biology, Faculty of Computer Science, University of Vienna, 1090 Vienna, Austria
    \email{jakob.lykke.andersen@univie.ac.at}}
\affil[4]{Institute for Theoretical Chemistry, University of Vienna, 1090 Wien, Austria
	\email{xtof@tbi.univie.ac.at}}
\affil[5]{Bioinformatics Group, Department of Computer Science, and
Interdisciplinary Center for Bioinformatics, and German Centre for
Integrative Biodiversity Research (iDiv) Halle-Jena-Leipzig and Competence
Center for Scalable Data Services and Solutions Dresden-Leipzig and Leipzig
Research Center for Civilization Diseases, University of Leipzig, Leipzig
D-04107, Germany
	\email{stadler@bioinf.uni-leipzig.de}}
\affil[6]{Max Planck Institute for Mathematics in the Sciences, Leipzig D-04103, Germany}
\affil[7]{Fraunhofer Institute for Cell Therapy and Immunology, Leipzig D-04103, Germany}
\affil[8]{Center for non-coding RNA in Technology and Health, University of Copenhagen, Frederiksberg C DK-1870, Denmark}
\affil[9]{Santa Fe Institute, 1399 Hyde Park Rd, Santa Fe NM 87501, USA}
\affil[10]{Research Network Chemistry Meets Microbiology, University of Vienna, Wien A-1090, Austria}
\date{}

\begin{document}
\maketitle
\clearpage
\begin{abstract}
  We present an elaborate framework for formally modelling pathways in
  chemical reaction networks on a mechanistic level.
  Networks are modelled mathematically as directed multi-hypergraphs, with
  vertices corresponding to molecules and hyperedges to reactions.
  Pathways are modelled as integer hyperflows and we expand the network model by detailed
  routing constraints.  In contrast to the more traditional approaches like
  Flux Balance Analysis or Elementary Mode analysis we insist on
  integer-valued flows.  While this choice makes it necessary to solve
  possibly hard integer linear programs, it has the advantage that more
  detailed mechanistic questions can be formulated.
It is thus possible to query networks for general transformation motifs,
and to automatically enumerate optimal and near-optimal pathways.
Similarities and differences between our work and
  traditional approaches in metabolic network analysis are discussed in
  detail.  To demonstrate the applicability of the mathematical framework
  to real-life problems we first explore the design space of possible
  non-oxidative glycolysis pathways and show that recent manually designed
  pathways can be further optimised.  We then use a model of
  sugar chemistry to investigate pathways in the autocatalytic formose process.
  A graph transformation-based approach is used to
  automatically generate the reaction networks of interest.
\end{abstract}

\section{Introduction}\label{sec:introduction}
Chemical reactions are intrinsically many-to-many
relationships on molecules.  Chemical reaction networks thus are naturally
modelled as directed multi-hyper\-graphs~\cite{Zeigarnik:00a}, with vertices
being molecules and directed hyperedges being reactions. A wide variety of
methods has been devised to characterise the function of the networks in
terms of pathways. Nevertheless, the very notion of a \emph{pathway} has
remained difficult to define formally. While biochemical tradition
stipulates that certain combinations of enzyme-catalysed reactions
constitute pathways, competing mathematical approaches single out reaction
sets with specific properties that may serve as natural pathways in a
particular setting. Flux Balance Analysis (FBA)
\cite{Fell:86,Kauffman:03,Orth:10} is geared towards finding a single
pathway that is optimal in a certain sense, usually w.r.t.\ a biomass
function defined from experiments.  In contrast, Elementary Flux Modes
(EFM) \cite{Schuster:94} and Extremal Pathways (ExPa)
\cite{Klamt:02,Klamt:03} consider particular sets of pathways that
implicitly describe the complete solution space. An important formal
difference is that EFMs and ExPas have a mechanistic interpretation as
integer valued fluxes, while FBA pathways do not have this property.
Reversible reactions are in ExPa represented as separate pairs of mutually
inverse hyperedges, while this is not done in FBA and EFM which instead
allow for negative fluxes.  A general mechanistic model related
to EFM pathways introduces additional constraints to limit futile flux~%
\cite{Beasley:01012007}. Interpretations of chemical networks that
disregard the stoichiometry of the network have been popular since they are
amenable to efficient methods from network analysis (e.g., see
\cite{Planes:20092244}) but are much less expressive~\cite{Planes:01092008}.

We introduce here a versatile model for general mechanistic pathways, where
routing constraints can be introduced to limit futile branches.  Pathways
are modelled as \emph{integer}-valued hyperflows, which are the natural
generalization of flows on directed graphs to directed hypergraphs~%
\cite{Gallo:93}. Hyperflows are mathematically equivalent to chemical
fluxes. They have been studied in their own right for restricted classes of
hypergraphs \cite{Cambini:97,Gallo:98a} and provide a direct link to the
rich computer science literature (see e.g.,~\cite{ahuja}). Hyperflows can
be found with Linear Programming (LP), as in FBA. We will show, however,
that there are cases where it is important to insist on integer hyperflows.
Integer hyperflows can be found with Integer Linear Programming (ILP), a
technique that has been applied in previous pathway models
\cite{Planes:01092008,Beasley:01012007} and that we will also make
extensive use of. As we shall see, LP and ILP solutions can be vastly
different. It is well-known that LP can be solved in polynomial time
\cite{lpPoly1,lpPoly2}, while ILP is NP-hard in the general case
\cite{Karp:72}. Not surprisingly, the restricted problem of finding a
maximum integer hyperflow in a chemical reaction network is also NP-hard,
even for bi-molecular reactions (bounded degree
hyperedges)~\cite{andersen:12}. Nevertheless, modern ILP solvers readily
deal with most problem instances of practical relevance.  Specialised
(Mixed) ILP modification schemes have been devised to enumerate EFMs
\cite{Figueiredo:2009} and minimal reaction sets in an FBA setting
\cite{Jonnalagadda2014,Burgard2001,Rohl2017}.  Here we use a tree search
algorithm in combination with the underlying ILP solver to enumerate
alternative optimal and near-optimal pathways.

Pathways are only one way to capture the structure of a chemical network.
The theory of Chemical Organizations (CO) \cite{Kaleta:06,Centler:08}
considers higher-level properties of pathways and focusses on closure
properties of subnetworks. A closely related topic is the recognition of
catalytic and autocatalytic cycles. Autocatalysis forms the basis of
autopoietic systems, a particular type of organizational structure. In
prebiotic chemistry \cite{Ruiz-Mirazo:2014}, autocatalysis has long been
hypothesized to be one of the dominating mechanisms that eventually lead up
to the origin of life. Logically described by the simple scheme $(A)+X\to
n\,X +(B)$, autocatalysis may be instantiated by a wide variety of
distributed chemical schemes \cite{blackmond:2009,plasson:2011} akin to
autocatalytic set models \cite{Kauffman:86,Hordijk:13}. We will show that
the geometric interpretation of integer hyperflows also facilitates the
definition of algebraic criteria for autocatalyis, thereby allowing us to
identify autocatalytic cycles as the solutions of an ILP problem. To this
end we devise here an expanded network model that represents each vertex
(compound) as a bipartite digraph. This allows us to efficiently implement
additional topological constraints to handle futile cycles and to treat
catalytic and autocatalytic pathways that are not immediately meaningful in
the traditional FBA framework without mechanistic pathways.

From a chemical point of view the notion of a \emph{chemical transformation
  motif} (CTM) takes on a central role in this context. It refers to a
coherent subset of chemical reactions with a well-defined interface to the
remainder of the network and implements a coherent overall chemical
transformation. Metabolic pathways and subnetworks exhibiting overall
(auto)catalysis are particular examples of CTMs. Hence CTMs are formal
specification of a set of integer hyperflows in a chemical reaction
network, possibly in conjunction with additional constraints on the allowed
hyperflows. As an example we will investigate the many ways of
instantiating the transformation motif \emph{defined} by the conversion of
1 glucose to 3 acetylphosphates and the alternative implementations of the
well-known autocatalytic Formose process. 

The approach presented here combines aspects of previous pathway models. It
can be used for finding single pathways as well as for the targeted,
large-scale enumeration of pathways, both with respect to a given
optimisation criterion, but without the burden of characterising the
complete solution space.

\section{Model}
\label{sec:model}
In this section we develop a formal model for pathways in reaction
networks.  For conciseness, we entirely use the language of hypergraphs.
We first describe the basic pathway model and a simple notion of
autocatalysis, characterised by the signature of the overall reaction of a
pathway.  This model allows for misleading, or chemically
``uninteresting'', pathways.  An expanded model is therefore introduced to
narrow the space of pathways.

  Note that the core model aims at formalising the notion of a chemical
  pathway, and therefore does not by itself include any optimality
  criteria. This model forms the basis of a practical implementation
  in terms of integer linear programs (ILP), which we will describe in
  detail in the following section. Different optimality criteria can then
  be added to the model in the form of linear objective functions.

\subsection{Directed Multi-Hypergraphs} 
\begin{figure}
\centering
\incFig{%
\footnotesize
\begin{tikzpicture}
\node[hedge] (p) {};
\node[hnode] (g1) [below left of=p,xshift=-8,yshift=7] {\ch{g_1}};
\node[hnode] (g2) [above left of=p,xshift=-8,yshift=-7] {\ch{g_2}};
\node[hnode] (g3) [right of=p,xshift=4] {\ch{g_3}};
\multiedge{g1}{p}{1}
\multiedge{g2}{p}{1}
\multiedge{p}{g3}{1}
\end{tikzpicture}
\hfill
\begin{tikzpicture}
\node[hnode] (g4) {\ch{g_4}};
\node[hnode] (g5) [right of=g4,xshift=20] {\ch{g_5}};
\node[ellipse,minimum height=12, minimum width=17] (g1)
[below left of=g5,xshift=-0.7em,yshift=7] {\phantom{\ch{g_1}}};
\draw[edge] (g4) to node[auto] {} (g5);
\end{tikzpicture}
\hfill
\begin{tikzpicture}
\node[hnode] (g6) {\ch{g_6}};
\node[hedge] (p) [right of=g6,xshift=5] {};
\node[hnode] (g7) [right of=p,xshift=4] {\ch{g_7}};
\node[ellipse,minimum height=12, minimum width=17] (g1) 
[below left of=p,xshift=-0.7em,yshift=7] {\phantom{\ch{g_1}}};
\multiedge{g6}{p}{2}
\multiedge{p}{g7}{1}
\end{tikzpicture}%
} 
\caption[]{Visualization of an abstract reaction network 
consisting of the reactions 
\ch{g_1 + g_2 \carrow  g_3}, 
\ch{g_4 \carrow g_5}, and 
\ch{2\,g_6 \carrow g_7}.
Note that we use a simplified visualisation scheme for 1-to-1 reactions, without a box but simply with a direct arrow between the compounds.
}
\label{fig:networkViz}
\end{figure}
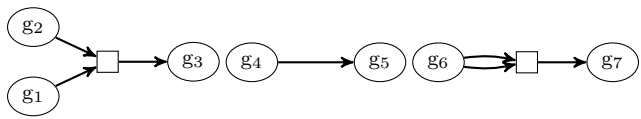

A directed multi-hypergraph $\mathcal{H} = (V, E)$ is an ordered
pair of a vertex set $V$ and a set of directed hyperedges $E$.  Each
edge $e\in E$ is itself a pair $(e^+, e^-)$ of multisets (hence
``multi-''hypergraph) of vertices $e^+\subseteq V$ and $e^-\subseteq V$,
called the tail and head of $e$. In the following we refer to
directed multi-hypergraphs simply as
hypergraphs. Fig.~\ref{fig:networkViz} gives an example of the
visualisation scheme we generally use throughout this contribution.

Since we will use both normal sets and multisets we introduce
the following notation for multisets.  When constructing a multiset we use
double curly brackets (i.e., $\mset{\dots}$) to distinguish them from
normal set constructors, $\{\dots\}$.  For a multiset $Q$ and some element
$q$ we use $m_q(Q)$ to denote the number of occurrences of $q$ in $Q$.  We
introduce a multi-membership operator, $\in_m$, used in iteration
contexts. E.g., $\sum_{q\in_m Q}1 = m_q(Q) = |\mset{q\in_m Q}|$.

\subsection{Integer Flows on Extended Hypergraphs}
\label{sec:basicModel}
Throughout, we assume that $\mathcal{H} = (V,E)$ is a directed
multi-hypergraph.  Syntactically we will use a superscripted plus~$^+$ to
refer to ``out''-related elements relative to vertices (e.g., out-edges),
and a superscripted minus $^-$ to refer to ``in''-related elements.

\begin{figure}
\centering
\removeStuff{%
\newcommand{\makeIOnode}[5][0.60]{{
	\newcommand{\Angle}{#3}
	\newcommand{\Anchor}{#2}
	\newcommand{\Dist}{#1}
	\newcommand{\InText}{#4}
	\newcommand{\OutText}{#5}
	\newcommand{\Offset}{15}
	\node[hnode, overlay, draw=none] (\Anchor-IO) at ($(\Anchor.\Angle)+(\Angle:\Dist)$) [anchor={\Angle-180}] {\vphantom{A}};
	\draw[edge] (\Anchor.\Angle+\Offset) to node[auto]{\OutText} (\Anchor-IO.\Angle-180-\Offset);
	\draw[edge] (\Anchor-IO.\Angle-180+\Offset) to node[auto]{\InText} (\Anchor.\Angle-\Offset);
}}
\footnotesize
\tikzset{
	spacer/.style={minimum size=2, inner sep=0},
	every matrix/.style={ampersand replacement=\&, row sep=4, column sep=16}
}%
}
\subcaptionbox{\label{fig:modelEx:network}}{
\incFig{%
\begin{tikzpicture}
\matrix{
\node[hnode] (F) {$F$};		\\
\node[hnode] (B) {$B$};	\& \node[hedge] (Be) {\phantom{1}};	\& \node[hnode] (C) {$C$};	\\
\node[spacer] {};		\\
\node[hnode] (A) {$A$};	\&\& \node[hedge] (Ce) {\phantom{1}};		\\
};

\draw[edge] (A) to [bend right=-10] node[auto] {} (B);
\draw[edge] (B) to [bend right=-10] node[auto] {} (A);

\multiedge{B}{Be}{1}\multiedge{F}{Be}{2}\multiedge{Be}{C}{1}
\multiedge{C}{Ce}{1}\multiedge{Ce}{A}{1}\multiedge{Ce}{B}{1}
\end{tikzpicture}
}
}%
\hspace{5em}
\subcaptionbox{\label{fig:modelEx:extended}}{
\incFig{%
\begin{tikzpicture}
\matrix{
\node[hnode] (F) {$F$};		\\
\node[hnode] (B) {$B$};	\& \node[hedge] (Be) {\phantom{1}};	\& \node[hnode] (C) {$C$};	\\
\node[spacer] {};		\\
\node[hnode] (A) {$A$};	\&\& \node[hedge] (Ce) {\phantom{1}};		\\
};
\makeIOnode{A}{180}{}{}
\makeIOnode{B}{180}{}{}
\makeIOnode{C}{0}{}{}
\makeIOnode{F}{180}{}{}

\draw[edge] (A) to [bend right=-10] node[auto] {} (B);
\draw[edge] (B) to [bend right=-10] node[auto] {} (A);

\multiedge{B}{Be}{1}\multiedge{F}{Be}{2}\multiedge{Be}{C}{1}
\multiedge{C}{Ce}{1}\multiedge{Ce}{A}{1}\multiedge{Ce}{B}{1}
\end{tikzpicture}
} 
}%
\caption[]{\subref{fig:modelEx:network} a small network,
  $\mathcal{H}$. \subref{fig:modelEx:extended} the extended network,
  $\overline{\mathcal{H}}$.
  Note that most of the input/output edges in the extended network will
  be constrained in the final formulation, and thus for many chemical
  networks many of these edges will effectively be removed to model
  specific interface conditions.
}
\label{fig:modelEx}
\end{figure}
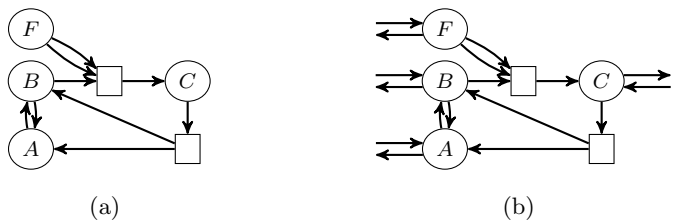

We need a mechanism to inject and extract molecules from the
network, i.e., what is called ``exchange reactions'' in metabolic networks.
We therefore define the \emph{extended hypergraph}
$\overline{\mathcal{H}}=(V, \overline{E})$ of $\mathcal{H}$ with 
$\overline{E} = E\cup E^- \cup E^+$ where 
$E^- = \{e^-_v = (\emptyset, \mset{v}) \mid v\in V\}$ and 
$E^+ = \{e^+_v = (\mset{v}, \emptyset) \mid v\in V\}$
designate the additional ``half-edges'' $e^-_v$ and $e^+_v$, for each $v\in V$.
An example is shown in Fig.~\ref{fig:modelEx}.

For a vertex $v\in V$ and an edge $e\in \overline{E}$ we use the multiplicity function for multisets to write
$m_v(e^-)$ for the number of occurrences of $v$ in the head of $e$ (and $m_v(e^+)$ for the tail).
We use $\delta^+(\cdot)$ and $\delta^-(\cdot)$ to denote a set of
incident out-edges and in-edges respectively, and thus use
$\delta^{+}_{\overline{E}}(v)$ as the set of out-edges from $v$, restricted
to the edge set $\overline{E}$, i.e., $\delta^{+}_{\overline{E}}(v)=\{e\in
\overline{E}\mid v\in e^{+}\}$.
Likewise, $\delta^{-}_{\overline{E}}(v)$
denotes the restricted set of incident in-edges of $v$.
\begin{definition}
  An integer hyperflow on $\overline{\mathcal{H}}$ is a function $f\colon
  \overline{E}\rightarrow \mathbb{N}_0$ satisfying, for each $v\in V$ the
  conservation constraint
\begin{equation}
\label{eq:masscons}
  \sum_{e\in \delta^+_{\overline{E}}(v)} m_v(e^+)f(e) - 
  \sum_{e\in \delta^-_{\overline{E}}(v)} m_v(e^-)f(e) = 0
\end{equation}%
That is, the sum of flow out of each vertex must be the same as the sum of flow into it.
We mostly speak of integer hyperflows, and will for brevity refer to
them simply as flows.
\end{definition}
  
In order to constrain the in- and out-flow to certain vertices we specify
a set of inputs (sources) $S\subseteq V$ and outputs (targets/sinks)
$T\subseteq V$. Thus
\begin{equation} 
  f(e^-_v) = 0 \quad\forall v\not\in S \quad\textrm{and}\quad
  f(e^+_v) = 0 \quad\forall v\not\in T  		\label{eq:ioConstraint}
\end{equation}
serve as additional constraints in an I/O-constrained extended hypergraph,
which is completely specified by the the triple $(\mathcal{H},S,T)$.

We adopt the notion of an \emph{overall flow} from the chemical
\emph{overall reaction} for a pathway, which is simply a convenient
notation for the I/O flow.  For a flow $f$ we syntactically write the
overall flow as
\begin{equation*}
\sum_{i=1}^{|V|} f(e^-_{v_i})\, v_i 
\carrow 
\sum_{i=1}^{|V|} f(e^+_{v_i})\, v_i 
\end{equation*}
However, as usual for chemistry we omit the terms with vanishing
coefficients.

Flows are non-negative by definition. It is therefore necessary to
model every reversible reaction by two separate edges $e=(e^+,e^-)$ and
$e'=(e^-,e^+)$.  This separation of the flow will later allow us to define
useful chemical constraints on the flow.

Finally, as in many typical flow problems, a capacity function
$u\colon\overline{E}\to\mathbb{N}_0$ can be added to limit the flow
from above, i.e., $f(e)\le u(e)$.  We will, however, not explicitly
make use of a capacity function in this contribution.

\subsection{Specialised Flows -- Overall (Auto)catalysis}
The nature of autocatalysis is that the product of a chemical reaction
catalyses its own formation. This interaction leads to an exponential time
behaviour in the growth characteristics of the product, as well as, to a
positive correlation of initial concentration and the reaction rate.
Autocatalytic reactions have been investigated for over a century and
instances of this behaviour have been found in a wide range of topologically
different chemical systems, which are based on a rich chemical setup (for a
recent review see \cite{Bissette:2013}).
Usually several reactions are organised in a cyclic network to achieve the proper topology for autocatalysis.

We here define a simple notion of both catalysis and autocatalysis in terms
of the I/O flow of a network.  As we only constrain the flow of the overall
reaction, we call this \emph{overall} (auto)catalysis.  Catalysis means in
chemical terms that a molecule, the catalyst, is first consumed by some
reaction and then regenerated by a subsequent reaction in such a way that
overall the catalyst is neither consumed nor produced. We therefore say
that a vertex $v\in V$ is \emph{overall catalytic} in a flow $f$ if and
only if (i) the input and output flows of $v$ are non-zero and (ii) the
input and output flows of $v$ are equal, i.e., iff
\begin{align}
0 &< f(e_v^-) = f(e_v^+) 		\label{eq:overallCatalysisConstraint}
\end{align}
Similarly, autocatalysis, refers to a situation where a molecule is
consumed in a reaction only to be (re)produced in subsequent reactions in
higher quantities than what has been consumed originally.  In terms of flow
we say a vertex $v\in V$ is \emph{overall autocatalytic} in a flow $f$ if
and only if
\begin{align}
0 &< f(e_v^-) < f(e_v^+) 		
\label{eq:overallAutocatalysisConstraint}
\end{align}
We extend the terminology to say that a flow $f$ is overall (auto)catalytic
if some vertex is overall (auto)catalytic in $f$.

\subsection{Chemically Simple Flow and Vertex Expansion}

Representing reversible reactions as pairs of irreversible ones gives
rise to trivial pathways consisting of edge $e$ and its inverse
  $e^{-1}$ with equal positive flow. Consider the example in
Fig.~\ref{fig:vertexExp:1}.
\begin{figure*}
\centering
\removeStuff{%
\tikzset{
	hnode/.style={hnodeNoDraw, draw},
	every matrix/.style={row sep=6, column sep=10, ampersand replacement=\&},
	node distance=0.5
}%
}
\footnotesize
\subcaptionbox{$f_1$\label{fig:vertexExp:1}}{%
\incFig{%
\begin{tikzpicture}
\matrix{
\&\& \node[hnode] (AB) {$\phantom{A}$};	\&\& \node[hnode] (C) {$C$};
	\&\&\& \node[hedge] (BCe) {1};	\\
\& \node[hedge] (Ae) {1};	\&\& \node[hedge] (Be) {1};	\&\&\&\&\& \node[hnode] (D) {$D$};	\\
\node[hnode] (A) {$A$};	\&\& \node[hnode] (BA) {$\phantom{A}$};	\&\& \node[hnode] (B) {$B$};
	\&\&\& \node[hedge] (De) {1};	\\
};
\pgfmathsetmacro{\xoff}{0}
\pgfmathsetmacro{\yoff}{0}
\multiedge{A}{Ae}{1}\multiedge{Ae}{AB}{1}\multiedge{AB}{Be}{1}\multiedge{Be}{B}{1}
\multiedge{Be}{BA}{1}\multiedge{BA}{Ae}{1}
\node[overlay, hnodeNoDraw] (Ain) [below=of A] {\phantom{A}};
\draw[edge] (Ain) to node[auto]{1} (A);
\node[overlay, hnodeNoDraw] (Bio) [below=of B] {\phantom{B}};
\draw[edge] (B.-74) to node[auto]{2} (Bio.74);
\draw[edge] (Bio.106) to node[auto]{1} (B.-106);
\draw[edge] (B) to [bend right=-10] node[auto] {1} (C);
\draw[edge] (C) to [bend right=-10] node[auto] {1} (B);
\multiedge{B}{BCe}{1}\multiedge{C}{BCe}{1}\multiedge{BCe}{D}{1}
\multiedge{D}{De}{1}\multiedge{De}{B}{1}\multiedge{De}{C}{1}
\end{tikzpicture}%
} 
}%
\hfill
\subcaptionbox{$f_2$\label{fig:vertexExp:2}}{%
\incFig{%
\begin{tikzpicture}
\matrix{
\&\& \node[hnode] (AB) {$\phantom{A}$};	\&\& \node[hnode] (C) {$C$};		\\
\& \node[hedge] (Ae) {1};	\&\& \node[hedge] (Be) {1};	\\
\node[hnode] (A) {$A$};	\&\& \node[hnode] (BA) {$\phantom{A}$};	\&\& \node[hnode] (B) {$B$};	\\
};
\pgfmathsetmacro{\xoff}{0}
\pgfmathsetmacro{\yoff}{0}
\multiedge{A}{Ae}{1}\multiedge{Ae}{AB}{1}\multiedge{AB}{Be}{1}\multiedge{Be}{B}{1}
\multiedge{Be}{BA}{1}\multiedge{BA}{Ae}{1}
\node[overlay, hnodeNoDraw] (Ain) [below=of A] {\phantom{A}};
\draw[edge] (Ain) to node[auto]{1} (A);
\node[overlay, hnodeNoDraw] (Bio) [below=of B] {\phantom{B}};
\draw[edge] (B.-74) to node[auto]{2} (Bio.74);
\draw[edge] (Bio.106) to node[auto]{1} (B.-106);
\draw[edge] (B) to [bend right=-10] node[auto] {1} (C);
\draw[edge] (C) to [bend right=-10] node[auto] {1} (B);
\end{tikzpicture}%
} 
}%
\hfill
\subcaptionbox{$f_3$\label{fig:vertexExp:3}}{%
\incFig{%
\begin{tikzpicture}
\matrix{
\&\& \node[hnode] (AB) {$\phantom{A}$};	\\
\& \node[hedge] (Ae) {1};	\&\& \node[hedge] (Be) {1};	\\
\node[hnode] (A) {$A$};	\&\& \node[hnode] (BA) {$\phantom{A}$};	\&\& \node[hnode] (B) {$B$};	\\
};
\pgfmathsetmacro{\xoff}{0}
\pgfmathsetmacro{\yoff}{0}
\multiedge{A}{Ae}{1}\multiedge{Ae}{AB}{1}\multiedge{AB}{Be}{1}\multiedge{Be}{B}{1}
\multiedge{Be}{BA}{1}\multiedge{BA}{Ae}{1}
\node[overlay, hnodeNoDraw] (Ain) [below=of A] {\phantom{A}};
\draw[edge] (Ain) to node[auto]{1} (A);
\node[overlay, hnodeNoDraw] (Bio) [below=of B] {\phantom{B}};
\draw[edge] (B) to node[auto]{1} (Bio);
\end{tikzpicture}%
} 
}
\caption[Simplification of flows]{
Simplification of a flow $f_1$ to an equivalent flow $f_3$, by removal of futile 2-edge sub-pathways.
\subref{fig:vertexExp:1} The molecule $D$ is created through $B + C\carrow D$,
	but can only be interpreted as being consumed through the reverse reaction.
\subref{fig:vertexExp:2} After removal of 1 flow from the reactions $B + C\carrowBoth D$,
	the molecule $C$ now participates in a futile 2-edge flow.
\subref{fig:vertexExp:3} Removing 1 flow from $B\carrowBoth C$ and the I/O edges $\emptyset\carrowBoth B$, we arrive at the simplest flow.
}
\end{figure*}
Here we see three pairs of reversible reactions with positive flow: $B + C
\carrowBoth D$, $B \carrowBoth C$, and the I/O reactions $\emptyset
\carrowBoth B$.  However, we can argue that this flow is not ``simple'' in
the sense that there is no interpretation of the flow without a futile
conversion of matter. This problem has of course been recognized in the
literature. Additional ILP constraints were used in
\cite{Planes:01092008,Beasley:01012007} to prune such cases. However, this
approach seems quite difficult to control in practice.
Disallowing all cases of flow on both a reaction and its reverse as in \cite{Beasley:01012007}, 
for example, turns out to be too restrictive.
Another approach is to simply cancel out flow as a post-processing step \cite{Kaleta01102009}.
In the following we illustrate that certain seemingly futile flows
  should be allowed. To facilitate precise constraints we therefore advocate a refinement of the network model itself.

In the pathway a single copy of $D$ is created, through the reaction $B + C\carrow D$,
and it can only be routed into a single reaction, $D\carrow B + C$.
The sub-pathway $B + C\carrow D\carrow B + C$ is thus a futile 2-edge branch that we can simplify away, yielding the equivalent flow in Fig.~\ref{fig:vertexExp:2}.
The same reasoning can now be applied to $C$, and subsequently $B$, resulting in the flow depicted in Fig.~\ref{fig:vertexExp:3}.

The original flow in Fig.~\ref{fig:vertexExp:1} fulfils the requirement for overall autocatalysis in vertex $B$ (Eq.~\eqref{eq:overallAutocatalysisConstraint}),
but clearly the in-flow of 1\,$B$ is not involved in the extra production of $B$,
which goes against the idea of autocatalysis.
The simplified flow, Fig.~\ref{fig:vertexExp:3}, is however not overall autocatalytic,
and it is therefore desirable to constrain the model such that futile 2-edge flows are not possible,
in order to further approach a precise characterisation of autocatalysis.

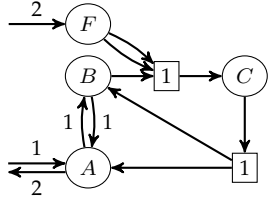
\begin{figure}
\centering
\incFig{%
\footnotesize
\begin{tikzpicture}[node distance=0.75]
\matrix[ampersand replacement=\&, row sep=4, column sep=16] {
\node[hnode] (F) {$F$};		\\
\node[hnode] (B) {$B$};	\& \node[hedge] (Be) {1};	\& \node[hnode] (C) {$C$};	\\
\node[hedgeNoDraw] {\phantom{1}};		\\
\node[hnode] (A) {$A$};	\&\& \node[hedge] (Ce) {1};		\\ 
};
\node[hnode, draw=none, overlay] (Aio) [left=of A] {\phantom{A}};
\draw[edge] (A.180+11) to node[auto]{2} (Aio.-11);
\draw[edge] (Aio.11) to node[auto]{1} (A.180-11);

\node[hnode, draw=none, overlay] (Fio) [left=of F] {\phantom{F}};
\draw[edge] (Fio) to node[auto]{2} (F);

\draw[edge] (A) to [bend right=-10] node[auto] {1} (B);
\draw[edge] (B) to [bend right=-10] node[auto] {1} (A);

\multiedge{B}{Be}{1}\multiedge{F}{Be}{2}\multiedge{Be}{C}{1}
\multiedge{C}{Ce}{1}\multiedge{Ce}{A}{1}\multiedge{Ce}{B}{1}
\end{tikzpicture}%
} 
\caption{Example of a flow with meaningful flow on an edge and its inverse, in the network from Fig.~\ref{fig:modelEx:extended}. Only edges with non-zero flow are shown.}
\label{fig:modelEx:flow}
\end{figure}

From the shown example it is tempting to simply disallow flow on both
  an edge and its inverse.  This is the approach effectively used in
FBA-related methods, and also in flows on normal graphs \cite{ahuja,digraphs}.
Since flow on an edge and its inverse is
  inherently a part of the I/O specification for autocatalysis, some
  exceptions must be made for I/O flow.  However, as illustrated with
  the counterexample in Fig.~\ref{fig:modelEx:flow} even internal
  edges can have seemingly futile 2-cycles.
We can interpret this flow such that no flow is directly reversed:
{\setlength\multicolsep{1ex}%
\begin{multicols}{2}\noindent
\begin{enumerate}
\item $\emptyset \carrow A$
\item $A\carrow B$
\item $\emptyset\carrow F$ twice
\item $B + 2\,F\carrow C$
\item $C\carrow A + B$
\item $B\carrow A$
\item $A\carrow \emptyset$ twice
\end{enumerate}
\end{multicols}}\noindent
Since this interpretation has the flow on mutually inverse edges temporally separated,
it should not be excluded from the pathway model.

To facilitate fine-grained control of flow where directly reversed flow can be disallowed we expand the extended hypergraph into a larger network.
Each vertex is expanded into a subnetwork that represents the routing of flow internally in the expanded vertex.
Formally, for each $v\in V$
\begin{align}
	V_v^- &= \{u^-_{v, e} \mid \forall e\in \delta^-_{\overline{E}}(v)\}	\label{eq:expandedInVertices}\\
	V_v^+ &= \{u^+_{v, e} \mid \forall e\in \delta^+_{\overline{E}}(v)\}	\label{eq:expandedOutVertices}\\
	E_v   &= \left\{\left(\mset{u^-}, \mset{u^+}\right) \mid u^-\in V_v^-, u^+\in V_v^+\right\}	 \notag
\end{align}
That is, $v$ is replaced with a bipartite graph $(V_v^-\cup V_v^+, E_v)$
with the vertex partitions representing the in-edges and out-edges of $v$
respectively, and the edge set being the complete set of edges from the
in-partition to the out-partition. We say that $E_v$ is the set of
\emph{transit edges} of $v$.

The original edges are reconnected in the natural way; for each $e = (e^+,
e^-)\in \overline{E}$ the reconnected edge is 
$\widetilde{e}= (\widetilde{e}^+, \widetilde{e}^-)$ with	
$\widetilde{e}^- = \mset{u^-_{v, e}\mid v \in_m e^-}$ 
and
$\widetilde{e}^+ = \mset{u^+_{v, e}\mid v \in_m e^+}$.
We finally define the expanded hypergraph 
$\widetilde{\mathcal{H}} = (\widetilde{V}, \widetilde{E})$ as
\begin{equation*}
  \widetilde{V} = \bigcup_{v\in V} V_v^- \cup \bigcup_{v\in V} V_v^+
  \quad\textrm{and}\quad
  \widetilde{E} = \bigcup_{v\in V}E_v \cup
			\{\widetilde{e}\mid e\in \overline{E}\}
\end{equation*}
We expand the definition of a flow function to
$f\colon\widetilde{E}\rightarrow \mathbb{N}_0$ and pose the usual conservation
constraints, but on $\widetilde{\mathcal{H}}$: for all $v\in \widetilde{V}$
\begin{align}
  \sum_{e\in \delta^+_{\widetilde{E}}(v)} m_v(e^+)f(e) - 
  \sum_{e\in \delta^-_{\widetilde{E}}(v)} m_v(e^-)f(e) &= 0	 
  \label{eq:masscons:expanded}
\end{align}
The I/O constraints translates directly to the expanded network.  In
Sec.~\ref{sec:expandedProps} we formally describe the relationship between
flows on the extended and the expanded network.

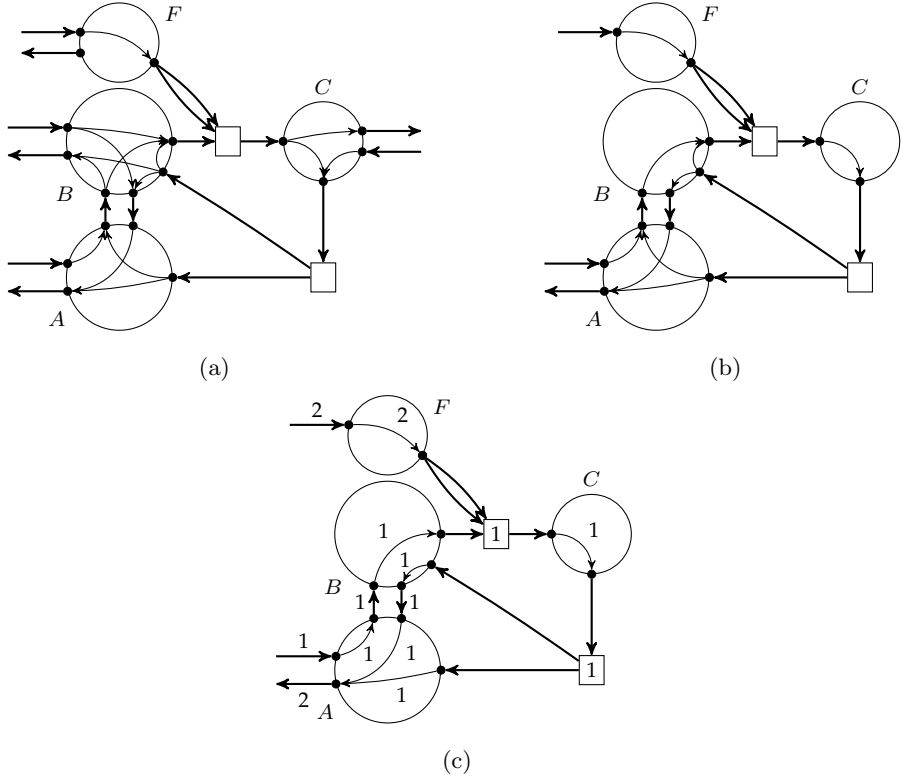
\begin{figure*}%
\centering
\subcaptionbox{\label{fig:modelEx:expanded:full}}{%
\incFig{%
\begin{tikzpicture}
\matrix{
\node[hxnodeF] (F) {};		\\
\node[hxnode, label=below left:$B$] (B) {};	\& \node[hedge] (Be) {\phantom{1}};	\& \node[hxnodeC] (C) {};	\\
\node[spacer] {};		\\
\node[hxnodeA] (A) {};	\&\& \node[hedge] (Ce) {\phantom{1}};		\\
};

\makeIOnodeL{A}{180}{}{}
\makeIOnodeL{B}{180}{}{}
\makeIOnode{C}{0}{}{}
\makeIOnode{F}{180}{}{}

\makeShortcutEdge{A}{105}{B}{-105}{}
\makeShortcutEdge{B}{-75}{A}{75}{}

\makeEdge{F/-30/10, F/-30/-10, B/0/0}{Be}{C/180/0}
\makeEdge{C/-90/0}{Ce}{A/0/0, B/-35/2}

\foreach \s/\t/\b in {A-in-IO/A-out-sc-B/30,	A-in-sc-B/A-out-IO/-40,	A-in-Ce/A-out-IO/-5, A-in-Ce/A-out-sc-B/-40,
		B-in-IO/B-out-sc-A/-40, B-in-IO/B-out-Be/-5,	B-in-sc-A/B-out-IO/40, B-in-sc-A/B-out-Be/-40,	B-in-Ce/B-out-sc-A/30, B-in-Ce/B-out-IO/5, B-in-Ce/B-out-Be/-60,
		C-in-Be/C-out-Ce/-45, C-in-Be/C-out-IO/-5,	C-in-IO/C-out-Ce/40,
		F-in-IO/F-out-Be/-25} {
	\draw[tedge] (t-\s) to [bend right=\b, looseness=1] (t-\t);
}
\end{tikzpicture}%
} 
}
\hfill
\subcaptionbox{\label{fig:modelEx:expanded:constrained}}{%
\incFig{%
\begin{tikzpicture}
\matrix{
\node[hxnodeF] (F) {};		\\
\node[hxnode, label=below left:$B$] (B) {};	\& \node[hedge] (Be) {\phantom{1}};	\& \node[hxnodeC] (C) {};	\\
\node[spacer] {};		\\
\node[hxnodeA] (A) {};	\&\& \node[hedge] (Ce) {\phantom{1}};		\\
};

\makeIOnodeL{A}{180}{}{}
\makeInode{F}{180}{}{}

\makeShortcutEdge{A}{105}{B}{-105}{}
\makeShortcutEdge{B}{-75}{A}{75}{}

\makeEdge{F/-30/10, F/-30/-10, B/0/0}{Be}{C/180/0}
\makeEdge{C/-90/0}{Ce}{A/0/0, B/-35/2}

\foreach \s/\t/\b in {A-in-IO/A-out-sc-B/30,	A-in-sc-B/A-out-IO/-40,	A-in-Ce/A-out-IO/-5, A-in-Ce/A-out-sc-B/-40,
		B-in-sc-A/B-out-Be/-40,	B-in-Ce/B-out-sc-A/30, B-in-Ce/B-out-Be/-60,
		C-in-Be/C-out-Ce/-45,
		F-in-IO/F-out-Be/-25} {
	\draw[tedge] (t-\s) to [bend right=\b, looseness=1] (t-\t);
}
\end{tikzpicture}%
} 
}
\hfill
\subcaptionbox{\label{fig:modelEx:expanded:flow}}{%
\incFig{%
\begin{tikzpicture}
\matrix{
\node[hxnodeF] (F) {};		\\
\node[hxnode, label=below left:$B$] (B) {};	\& \node[hedge] (Be) {1};	\& \node[hxnodeC] (C) {};	\\
\node[spacer] {};		\\
\node[hxnodeA] (A) {};	\&\& \node[hedge] (Ce) {1};		\\
};

\makeIOnodeL{A}{180}{1}{2}
\makeInode{F}{180}{2}

\makeShortcutEdge{A}{105}{B}{-105}{1}
\makeShortcutEdge{B}{-75}{A}{75}{1}

\makeEdge{F/-30/10, F/-30/-10, B/0/0}{Be}{C/180/0}
\makeEdge{C/-90/0}{Ce}{A/0/0, B/-35/2}

\foreach \s/\t/\b/\te in {A-in-Ce/A-out-IO/-5/1,
		B-in-sc-A/B-out-Be/-40/1,
		C-in-Be/C-out-Ce/-45/1,
		F-in-IO/F-out-Be/-25/2} {
	\draw[tedge] (t-\s) to [bend right=\b, looseness=1] node[auto] {\te} (t-\t);
}
\draw[tedge] (t-A-in-IO) to [bend right=30, looseness=1] node[auto, swap, xshift=-0.25em, yshift=0.25em] {1} (t-A-out-sc-B);
\draw[tedge] (t-A-in-sc-B) to [bend right=-40, looseness=1] node[auto, pos=0.15] {1} (t-A-out-IO);
\draw[tedge] (t-B-in-Ce) to [bend right=30, looseness=1] node[auto, swap, xshift=0.25em, yshift=-0.25em] {1} (t-B-out-sc-A);
\end{tikzpicture}%
} 
}
\caption[]{The network from Fig.~\ref{fig:modelEx:extended} expanded into
  $\widetilde{\mathcal{H}}$.
    The vertices of $\widetilde{\mathcal{H}}$ are the small filled circles,
  while the large circles, $A$, $B$, $C$ and $F$, only serves as visual
  grouping of the actual vertices.
  \subref{fig:modelEx:expanded:full} the
  expanded network without transit edges constrained in
  Eq.~\eqref{eq:reverseFlowConstraints}.
  \subref{fig:modelEx:expanded:constrained} the expanded network,
  simplified using the source set $S = \{A, F\}$ and sink set $T= \{A\}$.
  \subref{fig:modelEx:expanded:flow} the example flow from Fig.~\ref{fig:modelEx:flow} in the expanded
  network, with only edges with non-zero flow shown.  Note that no 2-cycles
  exist in this flow.
}%
\label{fig:modelEx:expanded}%
\label{fig:modelEx:flowExpanded}
\end{figure*}

Using the expanded network we can prevent flow from being directly reversed:
for each pair of mutually inverse edges $e = (e^+, e^-), e' = (e^-, e^+) \in \overline{E}$
we consider the transit edges they induce, i.e., the edges $(u^-_{v, e}, u^+_{v, e'})$ for all $v\in e^-$.
None of these edges may have flow through them.
A flow $f$ must thus satisfy
\begin{align}
f((u^-_{v, e}, u^+_{v, e'})) = 0 \qquad \forall v \in e^-    
\label{eq:reverseFlowConstraints}
\end{align}%
Fig.~\ref{fig:modelEx:expanded} shows the expanded version of the network
from Fig.~\ref{fig:modelEx:extended}, with these constraints in effect.

Note that the expansion of the network also opens the possibility of
forbidding other 2-sequences of edges, and in general the possibility of
posing constraints on the routing of flow internally in vertices.

When querying for chemical pathways with partially unknown I/O
specification we have found it useful to distinguish between reversible
reactions that are in the original network $\mathcal{H}$ and the reversible
I/O reactions.  That is, we may choose to not pose the above constraints on
transit edges $(u^-_{v, e}, u^+_{v, e'})$ when $e = e^-_v$ and $e' =
e^+_v$, thus allowing excess input-flow to be routed directly out of the
network again. Fig.~\ref{fig:modelEx:flow} shows a valid flow with a
meaningful 2-cycle. The expanded flow in
Fig.~\ref{fig:modelEx:flowExpanded} no longer has 2-cycles.

\begin{figure}
\centering
\subcaptionbox{\label{fig:strictCataProblem:nonStrict}}{%
\incFig{%
\begin{tikzpicture}
\commonPre
{2} 
{3} 
{2} 
\draw[tedge] (t-A-in-CtoA-1) to node[below, pos=0.5] {1} (t-A-out-ABtoC);
\end{tikzpicture}%
} 
}%
\hfill
\subcaptionbox{\label{fig:strictCataProblem:strict}} {%
\incFig{%
\begin{tikzpicture}
\commonPre
{1} 
{2} 
{1} 
\end{tikzpicture}%
} 
}%
\caption[]{A simplified network with an overall autocatalytic flow.
  \subref{fig:strictCataProblem:nonStrict} the vertex $A$ is both overall
  autocatalytic and an intermediate vertex in the flow.
  \subref{fig:strictCataProblem:strict} the same motif for overall
  autocatalysis, but without $A$ being an intermediate vertex.  }
\label{fig:strictCataProblem}
\end{figure}
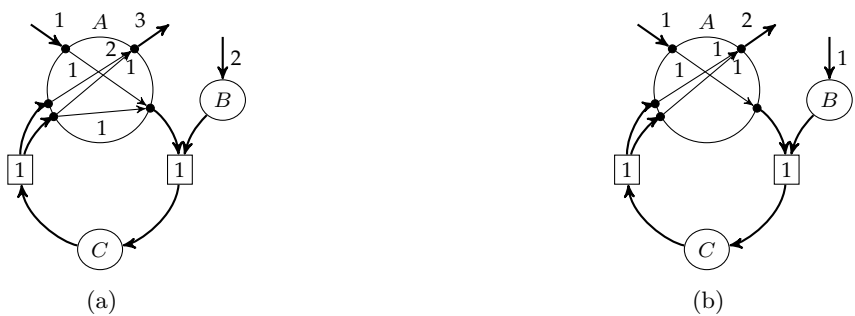

In Eq.~\eqref{eq:overallCatalysisConstraint}
and~\eqref{eq:overallAutocatalysisConstraint} we defined the I/O
constraints for overall catalysis and autocatalysis.  These constraints are
converted in the obvious manner to the expanded network
$\widetilde{\mathcal{H}}$.  However, the expanded network reveals another
possibility for somewhat misleading flows, exemplified in
Fig.~\ref{fig:strictCataProblem:nonStrict}.
Vertex $A$ is overall autocatalytic, but is also utilised as an intermediary molecule.
The same autocatalytic motif can be expressed by a simpler
flow, Fig.~\ref{fig:strictCataProblem:strict}.

In the interest of finding the simplest (auto)catalytic flows, we introduce
the following constraints.  Let $f$ be a flow and $v\in V$ a vertex
satisfying the I/O constraints for overall
catalysis~\eqref{eq:overallCatalysisConstraint} (resp.\ overall
autocatalysis~\eqref{eq:overallAutocatalysisConstraint}).  In the expanded
network $f$ must additionally satisfy the transit constraints (note that
$\delta_E^{\pm}(v)$ does not include the I/O edges):
\begin{align*}
f((u^+_{v, e'}, u^-_{v, e''})) &= 0	&	
\forall &e'\in \delta^-_E(v), e''\in \delta^+_E(v)
\end{align*}
That is, all transit flow in an overall (auto)catalytic vertex must flow
either from the input edge $e^-_v$ or towards the output edge $e^+_v$.

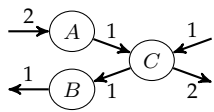
\begin{figure}
\centering
\incFig{%
\footnotesize
\begin{tikzpicture}
\node[hnode] (A) {$A$};
\node[hnode] (C) at ($(A) + (-20:4em)$) {$C$};
\node[hnode] (B) at ($(C) + (180+20:4em)$) {$B$};
\draw[edge] (A) to node[auto,above] {1} (C);
\draw[edge] (C) to node[auto,below] {1} (B);
\draw[edge] ($(A) + (-3em, 0)$) to node[above]{2} (A);
\draw[edge] (B) to node[above]{1} ($(B) + (-3em, 0)$);
\draw[edge] ($(C) + (20:3em)$) to node[auto,above]{1} (C);
\draw[edge] (C) to node[auto,below]{2} ($(C) + (-20:3em)$);
\end{tikzpicture}%
} 
\caption{The vertex $C$ is overall autocatalytic, but not autocatalytic in the chemical sense.}
\label{fig:exclusiveExample}
\end{figure}

If a compound is overall autocatalytic, it merely means that; if it is
available then even more can be produced.  However, this does not mean that
it cannot be produced solely by the other input compounds.  Solutions can
therefore be found that may be surprising.  One such solution is
illustrated in Fig.~\ref{fig:exclusiveExample}.
As a variant of our definition of overall autocatalysis we define
that a vertex $v$, is \emph{exclusively overall autocatalytic} if and only
if it is overall autocatalytic and is not \emph{trivially reachable} from
the other input vertices, $S\backslash\{v\}$.  A vertex $v$ is trivially
reachable from a vertex set $S'$ if it can be marked during a simple
breadth-first marking of the hypergraph $\mathcal{H} = (V, E)$.  For
completeness, the pseudocode is shown in Algorithm~\ref{alg:bfsMarking}.
\begin{algorithm}
\footnotesize
\SetKwInOut{Input}{Input}
\SetKwInOut{Output}{Output}
\Input{A directed (multi-)hypergraph $\mathcal{H} = (V, E)$.}
\Input{A set of starting vertices $S'\subseteq V$.}
\Output{A marked subset of the vertices.}
\lForEach{$v\in S'$}{mark $v$\;}
\While{no more hyperedges can be marked}{
	\ForEach{$(e^+, e^-)\in E$}{
		\If{all $v \in e^+$ are marked} {
			mark $e$\;
			\lForEach{$v\in e^-$}{mark $v$\;}
		}
	}
}
\caption{Breadth-first marking of a hypergraph.}
\label{alg:bfsMarking}
\end{algorithm}

Note that breadth-first marking of hypergraphs, and variations thereof, has
in the literature also been referred to as finding \emph{scopes} of
molecules 
\cite{scope1}.  
Breadth-first marking has in those studies been used
alone to analyse metabolic networks, and define set-theoretical notions of
pathways and later of autocatalysis \cite{kun:08}.  The methods thus do
not have focus on the underlying mechanism of the pathways, which is our aim in this contribution.

\subsection{Properties of the Expanded Hypergraph}
\label{sec:expandedProps}

The expansion of the networks obviously changes the size of the underlying
model, and it is therefore necessary to investigate how large the expanded
network can get, in order to bound the complexity of algorithms.  In this
section we only state the results. The proofs can be found in the
appendix.

For a directed multi-hypergraph $\mathcal{H} = (V, E)$, let the size of it
be denoted by $\mathrm{size}(\mathcal{H}) = |V| + |E| + \sum_{e\in E}(|e^+|
+ |e^-|)$.  Note that if the hypergraph is represented by a bipartite normal
graph, then this size corresponds to the number of vertices and edges.

\begin{proposition}
  The size of the extended network and the expanded network is polynomial
  in the size of the original network.
\end{proposition}

\begin{proposition}
  A feasible flow $f\colon \widetilde{E}\rightarrow \mathbb{N}_0$ on
  $\widetilde{\mathcal{H}}$ can be converted into an equivalent feasible
  flow $g\colon \overline{E}\rightarrow \mathbb{N}_0$ in
  $\overline{\mathcal{H}}$, with: $g(e) = f(\widetilde{e})$, for all $e\in
  \overline{E}$.
\end{proposition}

\begin{proposition}
  Let $f\colon \overline{E}\rightarrow \mathbb{N}_0$ be a feasible flow on
  $\overline{\mathcal{H}}$.  It can then be decided in polynomial time, in
  the size of $\mathcal{H}$, if a feasible flow $g\colon
  \widetilde{E}\rightarrow \mathbb{N}_0$ in $\widetilde{\mathcal{H}}$
  exists such that $g(\widetilde{e}) = f(e)$ for all $e\in \widetilde{E}$.
  If it exists it can be computed in polynomial time.
\end{proposition}

The problem of finding a pathway with maximum production of specific
compounds is known to be NP-hard, even for networks with bounded degree
reactions \cite{andersen:12}.  The last proposition underlines that the
expansion of the network does not drastically increase the complexity of
finding pathways, but the proof of it also shows a potentially practical algorithmic
approach to working with flows in the expanded network.
In Sec.~\ref{sec:implementation} we will however show an approach to
directly find the flows in the expanded network using ILP.

\subsection{Comparison to Existing Methods}
\label{sec:fbaComp}
The basic pathway model described in Sec.~\ref{sec:basicModel} is quite
similar to the formalism used in FBA, EFM and ExPa, with the latter two
methods primarily aiming to categorise specific classes of pathways
\cite{papin2004comparison}. The basic model is also similar to the model
presented in \cite{Beasley:01012007}, although the specification of allowed
I/O flow is phrased differently.  We briefly recast FBA in terms of
hypergraphs as the underlying model of reaction networks to clarify the
similarities but also the differences with our present approach.  The
mathematical development of FBA, EFM, and ExPa is based upon the concepts
of the \emph{stoichiometric matrix} and \emph{flux vectors}.  These are
analogous to a directed multi-hypergraph and non-integer hyperflows.  Let
$(\mathcal{H}, S, T)$ denote an I/O-constrained reaction network as defined
in Sec.~\ref{sec:basicModel}, with $\mathcal{H} = (V, E)$.  The network can
be represented as two matrices, an out-incidence matrix $\mathbf{S}^+$ and
an in-incidence matrix $\mathbf{S}^-$, both in the domain
$\mathbb{N}_0^{|V|\times |E|}$.  That is, each row represents molecules and
each column represents reactions.  Let vertices and hyperedges have some
arbitrary total order, $V = \{v_1, \dots, v_{|V|}\}$ and $E = \{e_1, \dots,
e_{|E|}\}$.  Then for each pair of vertices and reactions, $v_i, e_j$, the
matrices are defined as $\mathbf{S}^+_{i, j} = m_{v_i}(e_j^+)$ and
$\mathbf{S}^-_{i, j} = m_{v_i}(e_j^-)$.  Thus the columns of $\mathbf{S}^+$
represents the tail-multiset of each hyperedge, likewise for $\mathbf{S}^-$
and the head-multisets.  The actual stoichiometric matrix is defined as
$\mathbf{S} = \mathbf{S}^- - \mathbf{S}^+$, which in chemical terms is the
change of the number of each molecule that each reaction induces.  The
stoichiometric matrix describes both the proper chemical reactions and
transport of material from and to the outside, equivalent to the extension
of a hypergraph $\mathcal{H}$ to an extended hypergraph
$\overline{\mathcal{H}}$. 

The stoichiometric matrix $\mathbf{S}$ completely describes the original
reaction network, and thus is equivalent to the I/O-constrained reaction
network $(\mathcal{H}, S, T)$ if and only if all hyperedges have disjoint
head and tail. All direct catalysts, however, are cancelled out in the
stoichiometric matrix, hence the equivalence fails whenever there are
reaction hyperedges with $e^+\cap e^- \neq \emptyset$. This somewhat limits
the scope of FBA. Although it is possible in principle to replace reactions
with direct catalysts by a sequences of intermediate reactions that consume
and regenerate the catalyst, the resulting FBA network is no longer
equivalent to the original one and allows the drainage of
intermediates. This alters flux solutions and may be undesirable, e.g.,
when modelling the concerted action of enzyme complexes. Inspired by natural
biosynthetic pathways industrial biocatalysis research
\cite{Garcia-Junceda:2015,Choi:2015} currently intensively investigates
multi-enzyme cascade reactions, i.e., the combination of several enzyme
reactions in concurrent one-pot processes \cite{Ricca:2011}, because of
their prospect towards a ``greener'' and more sustainable chemical
future. Intermediates from these cascades cannot be accessed as substrates
by reactions outside the cascade; hence they require special treatment when
represented explicitly. 

A \emph{flux vector} $f\in\mathbb{R}^{|E|}$ models a pathway, and must
satisfy the usual conservation constraint, $\mathbf{S}\cdot f = 0$
(cmp.~Eq.~\eqref{eq:masscons}). Reversible reactions are modelled in one of two
ways:\\
(i) Combined: reversible reactions are modelled as a single reaction, but
  with the flow/flux allowed to be negative.  The flow/flux of irreversible
  reactions is non-negative.  This is the approach
  followed when finding EFMs \cite{Schuster:94}.\\
(ii) Separate: reversible reactions are modelled as two inverse reactions,
  and the flow/flux on all reactions must be non-negative.  This is the
  approached followed when finding ExPas \cite{schilling2000theory}.  We
  also follow this approach in our contribution both for mathematical
  simplicity and because it allows us to make use of the enhanced modelling
  capabilities offered by the expanded network.
\par
The extension of the stoichiometric matrix $\mathbf{S}$ to incorporate I/O
reactions can also be implemented using both the ``combined'' and the
``separate'' way of handling reversible reactions. The I/O constraints from
Eq.~\eqref{eq:ioConstraint}, specified by $S$ and $T$, translate naturally
to the corresponding constraints on the extended flux vector.

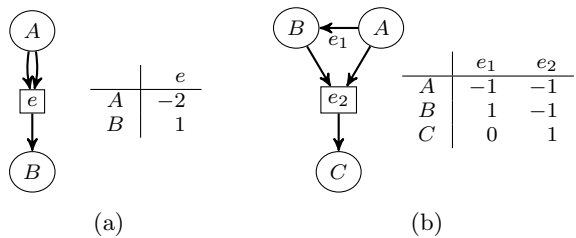
\begin{figure}
\centering
\removeStuff{%
\footnotesize
\tikzset{node distance=0.50}%
}%
\subcaptionbox{\label{fig:tumMulti}}{%
\incFig{%
\begin{tikzpicture}
\node[hnode] (A) {$A$};
\node[hedge, below=of A] (e) {$e$};
\node[hnode, below=of e] (B) {$B$};
\multiedge{A}{e}{2};
\draw[edge] (e) to (B);
\node[right=of e] {%
	\begin{tabular}{c|r}
		& $e$	\\\hline
	$A$	& $-2$	\\
	$B$	& $1$	\\
	\end{tabular}%
};
\end{tikzpicture}%
} 
}
\qquad
\subcaptionbox{\label{fig:tumSquare}}{%
\incFig{%
\begin{tikzpicture}
\node[hnode] (A) {$A$};
\node[hedge, below=of A, xshift=-2em] (e) {$e_2$};
\node[hnode, above=of e, xshift=-2em] (B) {$B$};
\node[hnode, below=of e] (C) {$C$};
\draw[edge] (A) to (e);
\draw[edge] (B) to (e);
\draw[edge] (e) to (C);
\draw[edge] (A) to node[auto] {$e_1$} (B);
\node[right=of e] {%
	\begin{tabular}{c|rr}
		& $e_1$	& $e_2$	\\\hline
	$A$	& $-1$	& $-1$	\\
	$B$	& $1$	& $-1$	\\
	$C$	& $0$	& $1$	\\
	\end{tabular}%
};
\end{tikzpicture}%
} 
}
\caption[]{Examples of reaction networks with not totally unimodular
  stoichiometric matrices. \subref{fig:tumMulti} all entries in a TU matrix
  must be $-1$, $0$, or $+1$. \subref{fig:tumSquare} the submatrix
  consisting of the top two rows has determinant $2$.  }
\label{fig:tum}
\end{figure}

With FBA we additionally define a linear objective function to find an
optimal flux vector, possibly with additional linear constraints
\cite{Savinell:92}.  As a flux vector is real-valued, and all the stated
constraints are linear, it can be found using linear programming (LP) in
polynomial time \cite{lpPoly1, lpPoly2}.  Herein lies a major difference to
the model presented in this contribution, where we require an
\emph{integer} hyperflow.  We can thus characterise the linear program from
FBA as the \emph{LP relaxation} of the basic pathway problem presented
here.

The LP relaxation of an ILP yields an integer solution only under special
conditions. The best known sufficient condition is that the matrix of
constraint coefficients is totally unimodular (TU), i.e., when all its
square submatrices have determinants $-1$, $0$, or $+1$, and thus all
entries of the matrix are also $-1$, $0$, or $+1$. This is the case for
example for integer flows in graphs \cite{ahuja,digraphs}. As the simple
examples in Fig.~\ref{fig:tum} shows, this not true in general for
stoichiometric matrices and hence for hyperflows.

Even though total unimodularity is not a necessary condition, it is not too
difficult to construct reaction networks with linear optimisation problems
where the integer problem and the LP relaxation have drastically different
optimal solutions. As an example consider the carbon rearrangement network
described later, in Sec.~\ref{sec:nog}, and the question: given 1 xylulose
5-phosphate (\ch{X5P}) and an arbitrary amount of phosphate ($P_i$), find
a pathway that maximises the production of acetylphosphate (\ch{AcP}).
As \ch{X5P} contains 5 carbon atoms and AcP contains 2, it is clear that
the maximum production from a single molecule must be at most 2 \ch{AcP}.
It turns out that the optimal integer solution produces just 1 \ch{AcP}, by the
single reaction \ch{P_i + X5P\carrow AcP + G3P + H_2O}.  However, the
optimal solution to the LP relaxation of the problem yields 2.5 \ch{AcP},
via the pathway described in Tab.~\ref{tab:fbaFail}.
\begin{table}
\centering
\incFig{%
\footnotesize
\colorlet{haxCol}{gray!20}
\newcommand{\colCellLeft}[1]{\multicolumn{1}{@{}>{\columncolor{haxCol}[0pt][\tabcolsep]}l}{#1}}
\newcommand{\colCell}[1]{\multicolumn{1}{>{\columncolor{haxCol}[0pt][\tabcolsep]}r@{}}{#1}}
\newcommand{\colCellRight}[1]{\multicolumn{1}{@{}>{\columncolor{haxCol}[0pt][0pt]}l@{}}{#1}}
\begin{tabular}{@{}lr@{}l@{}}
\toprule
Flow    & \multicolumn{2}{c}{Reaction}							\\
\midrule
1.0    & \ch{G3P + DHAP}    &    \ch{\carrow FBP}					\\
1.0    & \ch{G3P}    &    \ch{\carrow DHAP}						\\
0.5    & \ch{R5P}    &    \ch{\carrow X5P}							\\
0.5    & \ch{E4P + F6P}    &    \ch{\carrow G3P + S7P}				\\
1.5    & \colCell{\ch{P_i + X5P}}    &    \colCellRight{\carrow\ch{AcP + G3P + H_2O}}		\\
0.5    & \ch{P_i + F6P}    &    \ch{\carrow AcP + E4P + H_2O}		\\
0.5    & \ch{P_i + S7P}    &    \ch{\carrow AcP + R5P + H_2O}		\\
1.0    & \ch{FBP + H_2O}    &    \ch{\carrow P_i + F6P}				\\
\midrule
Overall    & \ch{X5P + 1.5\ P_i}    & \ch{\carrow 2.5\ AcP + 1.5\ H_2O}	\\
\bottomrule
\end{tabular}
}
\caption{A non-integer pathway with maximum production of AcP from 1 X5P.
The optimal integer solution consists of the shaded reaction with flow 1.
Relaxing the integrality constraint thus allows for a recycling pathway using \ch{G3P}
to produce 1.5 extra copies of \ch{AcP}.
See Sec.~\ref{sec:results} for a table of molecule abbreviations.}
\label{tab:fbaFail}
\end{table}
This non-integer pathway is incidentally a superposition of the
  optimal integer pathway and a recycling pathway for converting 1
\ch{G3P} into 1.5 \ch{AcP}.

A scaling of the non-integer flow with a factor 2 gives an integer
solution, and indeed our methods can also find this as well as many other
solutions implementing the same motif.  Since LP solutions with
integer-valued constraint matrices and objective functions with integer
coefficients are rational, it is mathematically always possibly to scale
the LP solution to integer values.  The actual numbers, however, may become
very large. Taking physiological constraints into account, the number of
available individual molecules may be small, as low as 100 copies
\cite{lownumbers1}, and even smaller for biological macromolecules
\cite{Fedoroff:02,Milo:15}.  Small numbers may have a profound influence on
chemical kinetics and can in many cases cause qualitative changes in the
behaviour of systems \cite{Gillespie:07,Kreyssig:14}.  It is an important
feature of the ILP framework to allow for directly expressing such
biological constraints. At the same time, it provides pathway solutions
that have a direct mechanistic interpretation.  On the other hand, in
  some cases where composite pseudo-reactions are used to model
  experimental results, as in the case of biomass as objective function,
  non-integer coefficients may appear in way that makes scaling difficult
  or undesirable. Such models, however, inherently are are approximations
  consistent with the LP relaxation. Mechanistic pathways are no longer 
  well-defined in such as setting.

In the previous example we maximised the production of a specific molecule,
and saw that the ILP solution have objective value 1 and the LP relaxation
have objective value 2.5.  The ratio between these values is known as the
\emph{integrality gap}, and it is known that this gap can scale with the
input instance.  For a simple example, consider the reaction networks
stemming from the polynomial-time reduction described in
the appendix, reducing the well-known
\textsc{Independent Set} problem to maximising the production of a single
molecule in a reaction network.
\begin{figure}
\centering
\subcaptionbox{\label{fig:independent:K3}} {%
\incFig{%
\footnotesize%
\begin{tikzpicture}
\node[hnode] (v-g) {$v_g$};
\node[hnode] (e-1-2) at (90:5em) {$v_{1,2}$};
\node[hnode] (e-1-3) at (-30:5em) {$v_{1,3}$};
\node[hnode] (e-2-3) at (-150:5em) {$v_{2,3}$};
\node[hedge,label=30:{$1 \mid\frac{1}{2}$}] (v-1) at (30:5em) {$e_1$};
\node[hedge,label=150:{$0 \mid\frac{1}{2}$}] (v-2) at (150:5em) {$e_2$};
\node[hedge,label=-90:{$0 \mid\frac{1}{2}$}] (v-3) at (-90:5em) {$e_3$};
\foreach \x in {1, 2, 3} \draw[edge] (v-\x) to (v-g);
\foreach \x/\y/\b in {1/2/17, 1/3/-17, 2/3/17} {
	\draw[edge] (e-\x-\y) to[bend left=\b] (v-\x);
	\draw[edge] (e-\x-\y) to[bend right=\b] (v-\y);
};
\draw[edge] (e-1-2) + (90:2.5em) to node[auto] {$1 \mid 1$} (e-1-2);
\draw[edge] (e-1-3) + (-30:2.5em) to node[auto, swap] {$1 \mid 1$} (e-1-3);
\draw[edge] (e-2-3) + (-150:2.5em) to node[auto] {$0 \mid 1$} (e-2-3);
\draw[edge] (v-g) to node[auto] {$1 \mid\frac{3}{2}$} (60:3.5em);
\end{tikzpicture}%
}
}\qquad
\subcaptionbox{\label{fig:independent:K4}} {%
\incFig{%
\footnotesize%
\begin{tikzpicture}
\matrix[ampersand replacement=\&, row sep=0.7em, column sep=1.2em, inner sep=0em, nodes={inner sep=0.3333em}] {
\node[hnode] (e-1-4) {$v_{1,4}$};	\& 	\& \node[hedge,label=90:{$1 \mid\frac{1}{2}$}] (v-1) {$e_1$};	\& 	\& \node[hnode] (e-1-2) {$v_{1,2}$};	\\
	\& \node[hnode] (e-1-3) {$v_{1,3}$};	\	\& 	\& 	\& 	\\
\node[hedge,label=180:{$0 \mid\frac{1}{2}$}] (v-4) {$e_4$};	\& 	\& \node[hnode] (v-g) {$v_g$};	\& 	\& \node[hedge,label=0:{$0 \mid\frac{1}{2}$}] (v-2) {$e_2$};	\\
	\& 	\& 	\& \node[hnode] (e-2-4) {$v_{2,4}$};	\	\& 	\\
\node[hnode] (e-3-4) {$v_{3,4}$};	\& 	\& \node[hedge,label=-90:{$0 \mid\frac{1}{2}$}] (v-3) {$e_3$};	\& 	\& \node[hnode] (e-2-3) {$v_{2,3}$};	\\
};
\foreach \x in {1, 2, 3, 4} \draw[edge] (v-\x) to (v-g);
\foreach \x/\y in {1/2, 1/4, 2/3, 3/4} {
	\draw[edge] (e-\x-\y) to (v-\x);
	\draw[edge] (e-\x-\y) to (v-\y);
};
\draw[edge] (e-2-4) to[bend right=15] (v-2);
\draw[edge] (e-2-4) to[out=180, in=-30] (v-4);
\draw[edge] (e-1-3) to[bend right=-15] (v-1);
\draw[edge] (e-1-3) to[out=-90, in=130] (v-3);

\draw[edge] (e-1-2) + (35:2.5em) to node[auto] {$1 \mid 1$} (e-1-2);
\draw[edge] (e-1-3) + (-180:3.5em) to node[auto] {$1 \mid 1$} (e-1-3);
\draw[edge] (e-1-4) + (145:2.5em) to node[auto, swap] {$1 \mid 1$} (e-1-4);
\draw[edge] (e-2-3) + (-35:2.5em) to node[auto, swap] {$0 \mid 1$} (e-2-3);
\draw[edge] (e-2-4) + (-90:2.5em) to node[auto] {$0 \mid 1$} (e-2-4);
\draw[edge] (e-3-4) + (-145:2.5em) to node[auto] {$0 \mid 1$} (e-3-4);
\draw[edge] (v-g) to node[above] {$1 \mid 2$} (20:4em);
\end{tikzpicture}%
}
}%
\caption[]{Reduction from the independent set problem to the problem of
  maximising output from a molecule in a reaction network, applied to the
  two graphs \subref{fig:independent:K3} $K_3$ and
  \subref{fig:independent:K4} $K_4$.  The hyperedges are annotated with
  both a feasible integer flow and a feasible non-integer flow, written as
  $\langle\text{integer}\rangle \mid \langle\text{non-integer}\rangle$.
  Allowing a maximum inflow of 1 to all vertices $v_{i,j}$ and maximising
  the outflow of $v_g$ corresponds to finding a maximum independent set in
  the original graph.  }
\label{fig:independent}
\end{figure}
Applying the reduction to complete graphs with $n$ vertices and formulating
the problem in terms of hyperflows, we obtain an integrality gap of
$\frac{n}{2}$: for integer flows we can use at most 1 reaction, thus giving
a maximum output of 1. When the integrality constraint is removed we can
let the flow be 0.5 on all reactions, giving an output of
$\frac{n}{2}$. The reaction networks for the complete graphs of size 3 and
4 are shown in Fig.~\ref{fig:independent}. This illustrates that the use of
the LP relaxation is not just a technical detail, but changes the nature of
the problem entirely.

\subsection{Implementation Using Integer Linear Programming}
\label{sec:implementation}
The ILP formulation characterising feasible flows is based on an expanded
hypergraph $\widetilde{\mathcal{H}} = (\widetilde{V}, \widetilde{E})$.  The
flow function is modelled by an integer variable $x_e$ for each edge $e\in
\widetilde E$, and by constraints for flow conservation.  The basic
constraints are thus $x_e\in \mathbb{N}_0$ for all $e\in \widetilde{E}$ and
$$\sum_{e\in \delta^+_{\widetilde{E}}(v)}m_v(e^+)\cdot x_e -
	\sum_{e\in \delta^-_{\widetilde{E}}(v)}m_v(e^-)\cdot x_e = 0$$
This definition is similar to an ILP formulation of a classical network
flow problem, but with important differences; $\widetilde{H}$ is a
hypergraph so an edge $e\in \widetilde{E}$ may be in both
$\delta^-_{\widetilde{E}}(u)$ and $\delta^-_{\widetilde{E}}(v)$ (or
$\delta^+_{\widetilde{E}}(u)$ and $\delta^+_{\widetilde{E}}(v)$) for $u\neq
v$.  Additionally, $\widetilde{H}$ is a \emph{multi}-hypergraph, and thus
the coefficients $m_v(e^+)$ and $m_v(e^-)$ are introduced, which may be
larger than $1$.

The basic formulation can be augmented with constraints, e.g., on the I/O
edges, and an objective function depending on the specific problem to be
solved.  Additionally, the constraints for chemical flows specified in
Eq.~\eqref{eq:reverseFlowConstraints} are added in the obvious way.
However, to reduce the size of the ILP we merge transit nodes where no incident transit edges have associated constraints,
i.e., those created from irreversible reactions.

In the following sections we describe constraints for finding overall
catalytic and autocatalytic flows.  For the formulation we will use $M$ to
denote a classical ``large enough'' constant, and as some parts of the
models for overall catalysis and autocatalysis are similar we will
first describe the formulation of these common parts.

In many cases of practical relevance not only the optimal solution is
relevant because alternative, near optimal solutions may be easier to find
and realize in an evolutionary context. A linear program may have an
uncountable number of optimal solutions as the variables are in the domain
of real numbers.  The solutions can, however, be described by enumerating
the, possibly exponentially many, corners of the optimal face of the
polyhedron defined by the linear program \cite{lpEnum2}. This has also been
applied to FBA \cite{fbaEnum}.  The restriction of ILP keeps the set of
candidate solution countable (even finite with flow capacity constraints),
and it becomes straight-forward to enumerate not only optimal solutions, but also near-optimal solutions,
see Sec.~\ref{sec:solEnum}.

In our definition of (auto)catalysis we require that if a vertex is
(auto)catalytic, then no flow can enter the vertex from the network and
exit the vertex to the network again.  Let $z_v$ be the indicator variable
for the vertex $v\in V$ being (auto)catalytic, then the requirement is
trivially enforced by the constraints $x_e \leq M\cdot (1- z_v)$ for all\\
$e = (u^-_{v, e'}, u^+_{v, e''}) 
\in V^-_v\times V^+_v : e' \neq e^-_v \wedge e'' \neq e^+_v$.

We model catalysis by introducing an indicator variable $z_v^c\in\{0, 1\}$
for each $v\in V$ indicating whether $v$ is catalytic or not.  Thus we can
enforce a solution to be catalytic by posing the constraint
$\sum_{v\in V}z_v^c \geq 1$.
The actual constraints for the indicator variables are obtained partially by
the constraints above on strictness of flow.  Below follows the last
requirement, Eq.~\ref{eq:overallCatalysisConstraint}, which is realised
through a set of auxiliary indicator variables, $z_v^0,z_v^<, z_v^>\in \{0,
1\}$
\begin{align*}
	x_v^- = x_v^+ = 0 \Leftrightarrow z_v^0 = 1
	&\equiv\left\{\begin{aligned}
		1 - z_v^0 &\leq x_v^- +x_v^+				\\
		M\cdot (1- z_v^0) &\geq x_v^- + x_v^+
	\end{aligned}\right.							\\
	x_v^- < x_v^+ \Leftrightarrow z_v^< = 1
	&\equiv\left\{\begin{aligned}
		x_v^- &< x_v^+ + M\cdot (1-z_v^<)		\\
		x_v^- &\geq x_v^+ - M\cdot x_v^<	
	\end{aligned}\right.							\\
	x_v^+ < x_v^- \Leftrightarrow z_v^> = 1
	&\equiv\left\{\begin{aligned}
		x_v^+ &< x_v^- + M\cdot (1-z_v^>)		\\
		x_v^+ &\geq x_v^- - M\cdot z_v^>
	\end{aligned}\right.							\\
	0 < x_v^- = x_v^+ \Leftrightarrow z_v^c = 1
	&\equiv\left\{\begin{aligned}
		z_v^c &\geq 1 - z_v^< - z_v^> - z_v^0		\\
		z_v^c & \leq 1 - z_v^0						\\
		z_v^c & \leq 1 - z_v^<						\\
		z_v^c & \leq 1 - z_v^>
	\end{aligned}\right.
\end{align*}

As for overall catalysis we model overall autocatalysis with a
set of indicator variables $z_v^a\in \{0, 1\}$ for all $v\in V$, and force
a solution to overall autocatalytic with the constraint
$\sum_{v\in V}z_v^a \geq 1$.

We use the constraints for strictness of flow and model the remaining
constraint, Eq.~\ref{eq:overallAutocatalysisConstraint}, using the
auxiliary variable set $z_v^-$, indicating $x_v^- > 0$:
\begin{align*}
	0 < x_v^- \Leftrightarrow z_v^- = 1
	&\equiv\left\{\begin{aligned}
		z_v^- &\leq x_v^ - 										\\
		M\cdot z_v^-&\geq x_v^-
	\end{aligned}\right.											\\
	0 < x^-_v < x_v^+ \Leftrightarrow z_v^a = 1
	\\&\hspace{-2em}\equiv\left\{\begin{aligned}
		z_v^a &\leq x_v^- 										\\
		x_v^- &< x_v^+ + M\cdot (1 -z_v^a)						\\
		M\cdot z_v^a + x_v^- &\geq x_v^+ - M\cdot (1 - z_v^-)	
	\end{aligned}\right.
\end{align*}

\subsection{Solution Enumeration}
\label{sec:solEnum}
A typical use of solvers for integer programs is to find a single optimal
solution.  However, from a chemical perspective we are also interested in
near-optimal solutions and in some cases even all solutions.  The structure
of our formulation additionally have influence on when two solutions are
considered different.  Often we might not consider two solutions different
if they only differ in the flow on the transit edges, i.e., those
introduced by the vertex expansion.  This makes it difficult to use
build-in features in solvers, such as the solution pool in IBM ILOG CPLEX,
to enumerate solutions.

For finding multiple solutions we therefore explore a search tree based on
the domain of the variables; each vertex in the tree represents a
restriction of the variable domains, with children representing more
constrained domains.  Note that this tree, in theory, is infinite as some
variable may have no upper bound.  In each vertex we use an ILP solver to
find an optimal solution for the sub-problem.  If the problem is infeasible
the sub-tree is pruned, otherwise a path to a leaf in the tree is
constructed to represent the solution found by the ILP solver.  The quality
of the found solution at the same time acts as a lower bound on the
objective function of the sub-tree (when minimising the function).
Vertices in the tree are explored in order of increasing lower bound. If a
different value of flow is not to be considered a difference in the
solution we simply do not consider the corresponding variables to be part
of the branching procedure.

\subsection{Software}
The pathway model is implemented as an extension of the larger software
package, \modName{} (\modAbbr) \cite{mod0.5, mod}, which can be downloaded
at \url{http://mod.imada.sdu.dk/}.  The software combines the pathway
analysis with methods for working with generative chemistries
\cite{Andersen:13a,Andersen:2014c}, including the algorithms for network
expansion \cite{dgStrat} also used in this contribution.  The tight
integration between the methods makes it a convenient tool to design
artificial chemistries, both high-level systems like DNA-templated
computing \cite{dna}, or hypothetical prebiotic chemistries
\cite{hcn,eschenmoser}.  Our implementation uses IBM ILOG CPLEX 12.5 for
solving integer linear programs.  An upcoming version of \modAbbr{} is in
preparation that will include the pathway model, and a preliminary version
is accessible at \url{http://mod.imada.sdu.dk/playground.html}, with
examples of usage. The integrality constraints and the implicit constraints
depending on this assumption can easily be disabled in our implementation.
In this mode the framework is no longer guaranteed to produce mechanistic
pathways.  Instead, it reproduces classical FBA. Solution
enumeration, as described above, is not well-defined in this setting.

\section{Application Scenarios}
\label{sec:results}

Here we illustrate the strength of our modelling framework by analysing two
specific chemical systems. We shall see that beyond well studied objectives
such as the minimization of the number of reactions used or maximizing the
yield of a target compound, the objective can also be augmented with
constraints for a chemical transformation pattern, for example overall
autocatalysis. Both chemistries are modelled using the graph grammar
approach described in \cite{mod,dgStrat,Andersen:13a,Andersen:2014c}.  A
visualisation of all rules can be found in the appendix, along
with tables of molecule name abbreviations.

\subsection{Carbon Rearrangement in the Non-oxidative Glycolysis Pathway}
\label{sec:nog}
Microbial synthesis of plant-derived secondary metabolites currently is a
hot topic in metabolic engineering \cite{Thodey:2014,Nielsen:2015}.
Impressive examples include the synthesis of the anti-malarial drug
precursor artemisinic acid \cite{Ro:2006}, and of (S)-reticuline
\cite{DeLoache:2015}, an important intermediate towards the
benzylisoquinoline alkaloids codein or morphine.  The idea of engineering
microbes to synthesize useful compounds also extends to the production of
fuel.  One approach is to couple the catabolic pathway known as glycolysis,
which decomposes glucose (a \ch{C_6} molecule) into acetyl coenzyme A units
(\ch{AcCoA}, a \ch{C_2} molecule), to a designed synthetic pathway,
condensing \ch{AcCoA} into the skeleton of the target molecule.  The
drawback of this approach is, however, that 2 of the 6 carbon atoms from
glucose are lost as \ch{CO_2} during normal oxidative glycolysis, which
pushes the yield of this pathway down to \SI{75}{\percent}, in terms of
atom economy. A lossy process for producing educts for the synthetic
biofuel producing pathway is naturally a bad idea from an economic perspective.

A recent study \cite{bogorad2013synthetic} attacks the aforementioned
problem by hand crafting a non-oxidative glycolysis (NOG) pathway, which
prevents the carbon atom loss of its natural counterpart. The general logic
of this designed pathway is to couple the splitting reaction that produces
the desired \ch{C_2} body and a \ch{C_4} body as putative wast,
to a carbon rearrangement network, which then recycles the \ch{C_4} body into molecules,
that can be fed back into the NOG as educts.
With this strategy NOG achieves a \SI{100}{\percent} carbon atom economy.
The architecture of metabolic networks in
general shows this kind of self-referential topology; every putative waste
molecule is recycled, making metabolic networks very different from
unevolved chemical reaction networks (e.g., atmosphere or geochemical networks),
which do not show this property.

The paper\cite{bogorad2013synthetic} discusses several sources of variation
for the structure of the NOG pathway.  First, the splitting reaction can be
performed by two types of phosphoketolase (PK) enzymes, differing only in
the their input sugar preference; either fructose (F) or xylulose (X).
Second, the carbon rearrangement network can go either via fructose
1,6-bisphosphate (\ch{FBP}, a \ch{C_5} sugar) or sedoheptulose
1,7-bisphosphate (\ch{SBP}, a \ch{C_7} sugar).  Three pathways are
  shown (Fig.~2, \cite{bogorad2013synthetic}) that exploits this freedom in
  the design of the NOG pathway motif.  In this section we illustrate that
many more equivalent solutions can be found automatically, using
  enumeration of mechanistic pathways.  In order to explore related
  reactions for which concrete enzymes may not yet exist, we use a generic
  model of the chemistry.

The molecules are encoded as graphs in the straight-forward manner, though without stereochemical information.
This implies that certain classes of molecules are represented as a single molecule, e.g., \ch{Ru5P} and \ch{X5P}.

We have modelled the generic transformation rules listed in Tab.~\ref{tab:nogRules}, and shown in the appendix.
\begin{table}
\centering
\footnotesize
\begin{tabular}{@{}lll@{}}
\toprule
Abbr.	& Name	& Description \\
\midrule
AL		& Aldolase				& A generic aldol addition.				\\
AlKe	& Aldose-Ketose 		& Aldehyde to ketone conversion.			\\
KeAl	& Ketose-Aldose 		& Ketone to aldehyde conversion.			\\
PHL		& Phosphohydrolase	& Use water to cleave off phosphate. 		\\
PK		& Phosphoketolase		& Break C-C-bond and add phosphate.	\\
TAL		& Transaldolase			& Move \ch{C_3}-end.					\\
TKL		& Transketolase			& Move \ch{C_2}-end.					\\
\bottomrule
\end{tabular}%
\caption{List of generic transformation rules for modelling the non-oxidative glycolysis chemistry.
The full details of the rules are shown in the appendix.
}
\label{tab:nogRules}
\end{table}
In \cite{bogorad2013synthetic} the use of phosphoketolase is associated with specific names for the specific reactions:
\begin{itemize}
\item[] XPK for \ch{X5P + P_i \carrow AcP + G3P}
\item[] FPK for \ch{F6P + P_i \carrow AcP + E4P}
\end{itemize}
We extend the naming scheme to cover educts with 7 and 8 carbons:
\begin{itemize}
\item[] SPK for \ch{S7P + P_i \carrow AcP + X5P}
\item[] OPK for \ch{C_8P + P_i \carrow AcP + G6P} 
\end{itemize}

To create the reaction network we use the starting molecules \ch{P_i}, AcP, G3P, DHAP, E4P, R5P, Ru5P, F6P, S7P, and FBP.

The network is created by iterating the application of all the
transformation rules listed in Tab.~\ref{tab:nogRules}, until no new
molecules are discovered.  This chemistry is theoretically infinite, so we
impose the restriction that no molecule with more than 8 carbon atoms may
be created.  This is a reasonable constraint, since even under
physiological conditions carbohydrates are inherently metastable compounds.
For carbohydrates larger than \ch{C_8} the decomposition reactions become a
dominating reaction channel.  Although alternatives (L-type PPP) or
extended reaction sequences via higher carbohydrates have been suggested,
their biochemical evidence has been questioned (see discussion in recent
review \cite{Stincone:2014}). The network generation therefore terminates,
and results in a network with 81 molecules and 414 reactions.
The ILP for finding pathways contains 12077 variables, of which 10477 are due to transit edges.
Without transit node merging the number of transit edges would have been 23805.

For the overall reaction \ch{F6P + 2\, P_i\carrow 3\, AcP + 2\, H_2O} we
have enumerated all solutions using at most 8 unique reactions and with at
most 11 reactions happening in total.  Additionally we disable the AL and
PK reactions with small educt molecules; those with less than 3 carbon
atoms each.  This is in agreement with the experimentally characterised
reversible ping-pong mechanism of these two enzymes \cite{Kleijn:2005}.
This results in 263 different pathways, which were computed in 5 minutes on a desktop computer with an \computerDesc.
In Tab.~\ref{tab:nogTable} we categorise the pathways w.r.t.\ 
(i) the number of unique reactions used, (ii) the number of reactions,
(iii) whether the only bisphosphate used is FBP, and (iv)
the histogram of different PK reactions (see the modelling above).
The table shows the number of solutions for each combination.
\begin{table}
\centering
\incFig{%
\footnotesize
\input{nogTable}%
}
\caption{Overview of number of NOG pathways. Categories marked with subscript $_a$, $_b$, and $_c$ refer to Fig.~2 of \cite{bogorad2013synthetic}, and we see that not only are there alternate solutions in the exact same categories, but in the case of $_a$ we even find a shorter pathway with the same properties.
Note that the left block is similar to the middle block of categories, but with 1 less reactions used.
This is due to a replacement of part of the pathway with a shorter pathway, see Tab.~\ref{tab:nogDiff}.
The pathway shown in Fig.~\ref{fig:nogLessA} is from the framed blue category,
and the pathway in Fig.~\ref{fig:nogShortest} is from the framed green category.
Their counterparts with the replacement from Tab.~\ref{tab:nogDiff} are the unframed blue and green numbers.
}
\label{tab:nogTable}
\end{table}
Interestingly it turns out that the solution space where FBP is the only bisphosphate used is quite similar to the space where other bisphosphates are allowed but with only 7 unique reactions.
The solutions are distributed in the same manner except for a 1-shift in the number of reactions.
There is a 1-to-1 mapping between these two sets of solutions such that the only difference in the pathway is the sub-pathways described in Tab.~\ref{tab:nogDiff}.
\begin{table}
\centering
\footnotesize
\subcaptionbox{\label{tab:nogDiffLong}}{%
\begin{tabular}[b]{@{}rcl@{}}
\ch{FBP + H_2O}	& \ch{\xrightarrow{PHL}}	& \ch{P_i + F6P}		\\
\ch{E4P + F6P}		& \ch{\xrightarrow{TAL}}	& \ch{G3P + S7P}	\\
\ch{G3P + DHAP}	& \ch{\xrightarrow{AL}}		& \ch{FBP}			\\
\midrule
\ch{DHAP + E4P + H_2O}	& \ch{\longrightarrow}	& \ch{S7P + P_i}		\\
\end{tabular}%
}\hfill
\subcaptionbox{\label{tab:nogDiffShort}}{%
\begin{tabular}[b]{@{}rcl@{}}
\ch{DHAP + E4P}		& \ch{\xrightarrow{AL}}		& \ch{C_7P_2}	\\
\ch{H_2O + SBP}	& \ch{\xrightarrow{PHL}}	& \ch{S7P + P_i}		\\
\midrule
\ch{DHAP + E4P + H_2O}	& \ch{\longrightarrow}	& \ch{S7P + P_i}			\\
\end{tabular}%
}
\caption[]{Two pathways with the same overall reaction.
  \subref{tab:nogDiffLong} a 3-reaction pathway using FBP as bisphosphate.
  This sub-pathway is highlighted in Fig.~\ref{fig:nogLessA}.
  \subref{tab:nogDiffShort} a 2-reaction pathway using a bisphosphate with 7 carbons.
  This sub-pathway is highlighted in Fig.~\ref{fig:nogShortest}.
}
\label{tab:nogDiff}
\end{table}

\removeStuff{%
\newcommand{\molFig}[1]{\includegraphics[scale=0.20]{figures2/mols/#1}}%
\tikzset{nogTrickEdge/.style={draw=blue}}%
}
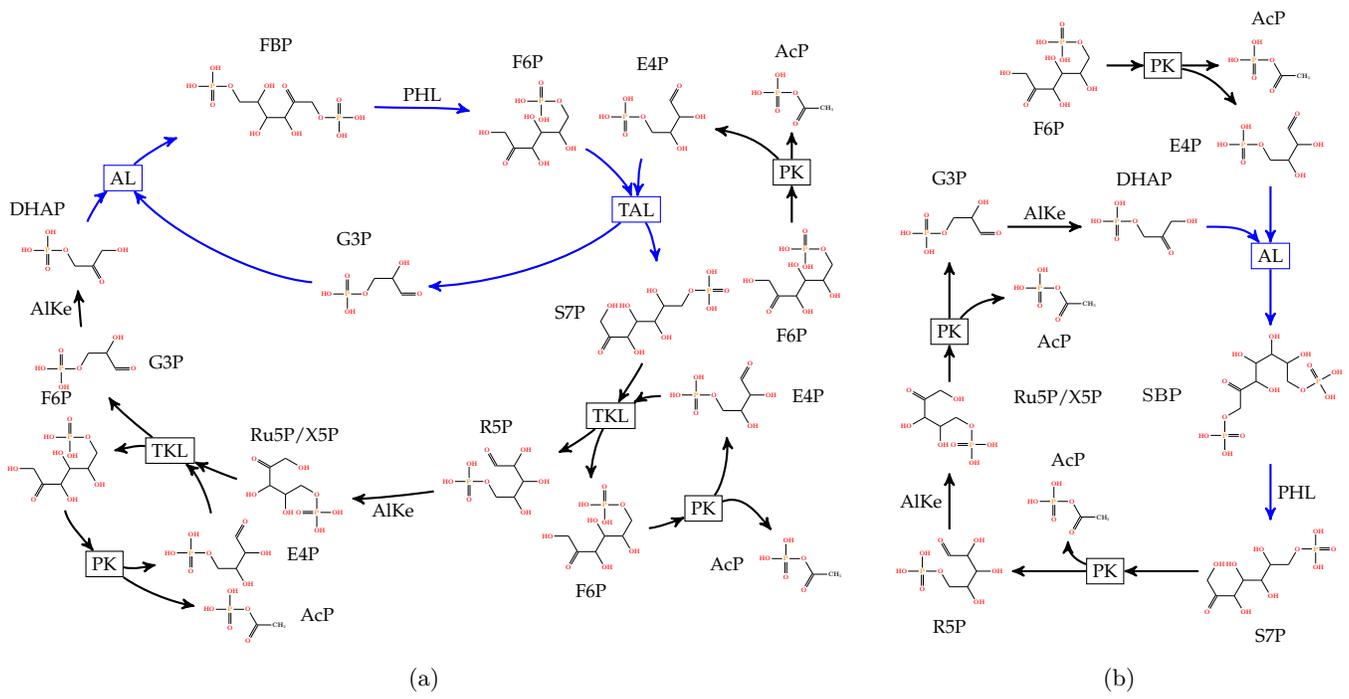
\begin{figure*}
\centering
\begin{adjustwidth}{-3cm}{-3cm}
\begin{center}
\subcaptionbox{\label{fig:nogLessA}}{%
\incFig{%
\scriptsize
\begin{tikzpicture}[nodeStyle/.style={hnodeNoDraw, rectangle, inner sep=2}]
\node[overlay, ellipse, minimum width=32em, minimum height=22em] (skeleton) {};

\node[hedge, nogTrickEdge, at=(skeleton.18)] (TAL) {TAL};
\node[nodeStyle, at=(skeleton.-3), label=180:S7P] (S7P) {\molFig{S7P}};
\node[hedge, at=(skeleton.-26)] (TKL-S7P) {TKL};
\node[nodeStyle, at=(skeleton.-55), label=90:R5P] (R5P) {\molFig{R5P}};
\node[nodeStyle, at=(skeleton.-110), label=90:Ru5P/X5P] (X5P) {\molFig{X5P}};
\node[hedge, at=(skeleton.-142)] (TKL-X5P) {TKL};
\node[nodeStyle, at=(skeleton.-167), label=0:G3P] (G3P-lower) {\molFig{G3P}};
\node[nodeStyle, at=(skeleton.172), label=95:DHAP] (DHAP) {\molFig{DHAP}};
\node[hedge, nogTrickEdge, at=(skeleton.154)] (AL) {AL};
\node[nodeStyle, at=(skeleton.115), label=90:FBP] (FBP) {\molFig{FBP}};
\node[nodeStyle, at=(skeleton.47), label=90:F6P] (F6P-FBP) {\molFig{F6P}};

\node[nodeStyle, at=($(TAL) + (80:5em)$), label=90:E4P] (E4P-FBP) {\molFig{E4P}};
\path[] (TAL) -|
		node[hedge, name=PKL-FBP, xshift=7.5em, yshift=2em] {PK}
	(E4P-FBP);
\node[nodeStyle, below=2em of PKL-FBP] (F6P-input) {\molFig{F6P}};
\node[below, at=(F6P-input.south)] {F6P};
\node[nodeStyle, at=($(PKL-FBP) + (90:4em)$), label=90:AcP] (ACP-FBP) {\molFig{AcP}};

\path[] (TAL) to[out=-140, in=-50, looseness=0.8]
		node[nodeStyle, pos=0.50, name=G3P-upper] {\molFig{G3P}}
	(AL);
\node[anchor=-40, at=(G3P-upper.90)] {G3P};

\node[overlay, ellipse, minimum width=10em, minimum height=7em, rotate=45, anchor=90, at=(TKL-S7P)] (S7P-skeleton) {};
\node[nodeStyle, at=(S7P-skeleton.-180), label=-90:F6P] (F6P-S7P) {\molFig{F6P}};
\node[hedge, at=(S7P-skeleton.-90)] (PKL-S7P) {PK};
	\node[nodeStyle, at=($(PKL-S7P) + (-30:6em)$), label=-180:AcP] (ACP-S7P) {\molFig{AcP}};
\node[nodeStyle, at=(S7P-skeleton.0), label=0:E4P] (E4P-S7P) {\molFig{E4P}};

\node[overlay, ellipse, minimum width=10em, minimum height=7em, rotate= -30, anchor=90, at=(TKL-X5P)] (X5P-skeleton) {};
\node[nodeStyle, at=(X5P-skeleton.180), label=90:F6P] (F6P-X5P) {\molFig{F6P}};
\node[hedge, at=(X5P-skeleton.-90)] (PKL-X5P) {PK};
	\node[nodeStyle, at=($(PKL-X5P) + (-20:8em)$), label=0:AcP] (ACP-X5P) {\molFig{AcP}};
\node[nodeStyle, at=(X5P-skeleton.0), label=0:E4P] (E4P-X5P) {\molFig{E4P}};

\draw[edge] (R5P) to[out=-171, in=-3, looseness=0.5] node[below] {AlKe} (X5P);
\draw[edge] (G3P-lower) to node[left] {AlKe} (DHAP);
\draw[edge, nogTrickEdge] (FBP) to[out=-3, in=167, looseness=0.5] node[above] {PHL} (F6P-FBP);

\draw[edge, nogTrickEdge] (E4P-FBP) to[out=-105, in=85, looseness=1] (TAL);
\draw[edge, nogTrickEdge] (F6P-FBP) to[out=-23, in=110, looseness=0.6] (TAL);
\draw[edge, nogTrickEdge] (TAL) to[out=-55, in=98, looseness=0.7] (S7P);
\draw[edge, nogTrickEdge] (TAL) to[out=-140, in=1, in looseness=0.5, out looseness=0.95] (G3P-upper);

\draw[edge] (S7P) to[out=-113, in=60, looseness=0.5] (TKL-S7P);
\draw[edge] (E4P-S7P) to[out=180, in=30, out looseness=0.7, in looseness=1] (TKL-S7P);
\draw[edge] (TKL-S7P) to[out=-140, in=25, looseness=0.7] (R5P);
\draw[edge] (TKL-S7P) to[out=-120, in=90, out looseness=0.8, in looseness=0.7] (F6P-S7P);

\draw[edge] (F6P-S7P) to[out=0, in=-155, looseness=0.5] (PKL-S7P);
\draw[edge] (PKL-S7P) to[out=50, in=-85, out looseness=0.5] (E4P-S7P);
\draw[edge] (PKL-S7P) to[out=20, in=130, looseness=1] (ACP-S7P);

\draw[edge] (X5P) to[out=164, in=-25, looseness=0.5] (TKL-X5P);
\draw[edge] (E4P-X5P) to[out=100, in=-40, looseness=0.7] (TKL-X5P);
\draw[edge] (TKL-X5P) to[out=143, in=-55, looseness=0.6] (G3P-lower);
\draw[edge] (TKL-X5P) to[out=-190, in=10, looseness=0.7] (F6P-X5P);

\draw[edge] (F6P-X5P) to[out=-80, in=130, looseness=0.7] (PKL-X5P);
\draw[edge] (PKL-X5P) to[out=-10, in=-170, looseness=1] (E4P-X5P);
\draw[edge] (PKL-X5P) to[out=-35, in=170, looseness=0.8] (ACP-X5P);

\draw[edge, nogTrickEdge] (G3P-upper) to[out=175, in=-50, looseness=0.7] (AL);
\draw[edge, nogTrickEdge] (DHAP) to[out=70, in=-140, looseness=0.5] (AL);
\draw[edge, nogTrickEdge] (AL) to[out=50, in=-160, looseness=0.5] (FBP);

\draw[edge] (F6P-input) to (PKL-FBP);
\draw[edge] (PKL-FBP) to[out=135, in=-10, looseness=1] (E4P-FBP);
\draw[edge] (PKL-FBP) to[out=90, in=-90, looseness=0.8] (ACP-FBP);
\end{tikzpicture}%
} 
}
\hfill
\subcaptionbox{\label{fig:nogShortest}}{%
\incFig{%
\scriptsize
\begin{tikzpicture}[nodeStyle/.style={hnodeNoDraw, rectangle, inner sep=2, node distance=0.75}]
\node[hedge, nogTrickEdge] (AL-C7P2) {AL};
\node[nodeStyle, below=of AL-C7P2, label=180:\ch{SBP}] (C7P2) {\molFig{C7P2_47}};
\node[nodeStyle, below=of C7P2, label=-90:S7P] (S7P) {\molFig{S7P}};
\node[hedge, left=4em of S7P] (PKL-S7P) {PK};
\node[nodeStyle, left=4em of PKL-S7P, label=-90:R5P] (R5P) {\molFig{R5P}};
	\node[nodeStyle, at=($(PKL-S7P) + (120:4em)$), label=90:AcP] (ACP-R5P) {\molFig{AcP}};
\node[nodeStyle, above=of R5P, label=15:Ru5P/X5P] (X5P) {\molFig{X5P}};
\node[hedge, above=2em of X5P] (PKL-X5P) {PK};
\node[nodeStyle, above=of PKL-X5P, label=90:G3P] (G3P) {\molFig{G3P}};
	\node[nodeStyle, at=($(PKL-X5P) + (20:6em)$), label=-90:AcP] (ACP-G3P) {\molFig{AcP}};
\node[nodeStyle, right=4em of G3P, label=90:DHAP] (DHAP) {\molFig{DHAP}};

\node[nodeStyle, above=of AL-C7P2, label=180:E4P] (E4P) {\molFig{E4P}};
\node[hedge, above left=2em of E4P] (PKL-F6P) {PK};
\node[nodeStyle, left=2em of PKL-F6P, label=-90:F6P] (F6P) {\molFig{F6P}};
	\node[nodeStyle, right=2em of PKL-F6P, label=90:AcP] (ACP-E4P) {\molFig{AcP}};

\draw[edge, nogTrickEdge] (E4P) to (AL-C7P2);	
\draw[edge, nogTrickEdge] (DHAP) to[out=0, in=130] (AL-C7P2);
\draw[edge, nogTrickEdge] (AL-C7P2) to (C7P2);

\draw[edge] (S7P) to (PKL-S7P);
\draw[edge] (PKL-S7P) to (R5P);
\draw[edge] (PKL-S7P) to[in=-90, out=160] (ACP-R5P);

\draw[edge] (X5P) to (PKL-X5P);
\draw[edge] (PKL-X5P) to (G3P);
\draw[edge] (PKL-X5P) to[in=-175, out=50] (ACP-G3P);

\draw[edge, nogTrickEdge] (C7P2) to node[auto] {PHL} (S7P);
\draw[edge] (R5P) to node[left] {AlKe} (X5P);
\draw[edge] (G3P) to node[above] {AlKe} (DHAP);

\draw[edge] (F6P) to (PKL-F6P);
\draw[edge] (PKL-F6P) to[out=-10, in=130] (E4P);
\draw[edge] (PKL-F6P) to (ACP-E4P);
\end{tikzpicture}%
} 
}
\end{center}
\end{adjustwidth}
\caption[]{%
	Two solution pathways for NOG.
	\subref{fig:nogLessA} A pathway similar to \cite[Fig.~2a]{bogorad2013synthetic}, but using fewer reactions.
	This solution category is the framed blue cell of Tab.~\ref{tab:nogTable}.
	The highlighted sub-pathway is the pathway from Tab.~\ref{tab:nogDiffLong}.
	\subref{fig:nogShortest} The shortest pathway found,
	denoted by the framed green cell in Tab.~\ref{tab:nogTable}.
	The highlighted sub-pathway is the pathway from Tab.~\ref{tab:nogDiffShort}.
}
\label{fig:nogSols}
\end{figure*}

In Fig.~\ref{fig:nogLessA} one of the solutions is illustrated in detail.
This solution has similar properties to the solution shown in Fig.~2a in
\cite{bogorad2013synthetic}: the phosphoketolase reactions all have F6P as
educt, and the only bisphosphate used is FBP.  However, this solution can
be regarded as being shorter as it uses fewer reactions, though the number
of unique reactions is the same.  Allowing other bisphosphates than FBP to
be used enables even shorter solutions to be found.
Fig.~\ref{fig:nogShortest} shows the shortest solution, which uses a
bisphosphate with 7 carbon atoms.  Its use of phosphoketolase is also
different, as it uses both XPK, FPK, and SPK.

The basic structure of the NOG solutions combine a productive splitting
reaction with a recycling network, which converts the ``waste'' produced
during the splitting reaction back into starting material using carbon
rearrangements. The major source of variability in the solutions stem from
the flexibility and combinatorial nature of the carbon rearrangement
chemistry. Our investigation nicely illustrates how vast the chemical
network design space is. Even if the reaction chemistry is restricted to
only a handful of enzyme functionalities, systematic exploration of the
network design space without computational approaches is inefficient and
many interesting solutions may be missed.

\subsection{Overall Autocatalysis in the Formose Process}
\label{sec:formose}

Carbohydrates \ch{(CH_2O)_n}, vulgo sugars, can formally be viewed as
polymers of formaldehyde \ch{CH_2O} building blocks. The chemical
reactivity of sugars is dominated by the two functional groups (1) carbonyl
group (\ch{C=O}) and (2) vicinal hydroxyl groups (\ch{HO-C-C-OH}), making
carbohydrates amenable to keto-enol tautomerization, aldol addition, and
retro-aldol fragmentation reactions. Due to this intrinsic reactivity,
sugars are potentially labile compounds, which readily isomerize into
complex mixtures under non-neutral conditions. The formose process,
described by Butlerov \cite{Butlerov:1861} is one of the scarce examples of
an autocatalytic reaction network, which generates complex mixtures of
sugars from an aqueous formaldehyde solution under high-pH conditions (for
a recent review on autocatalysis see \cite{Bissette:2013}). The formose
process has intensively been studied (for a recent review see
\cite{Delidovich:2014}) for its potential to produce biologically
significant carbohydrates from formaldehyde under prebiotic conditions
\cite{Benner:2010}. The time-concentration behaviour of formaldehyde
consumption during the formose process shows a linear lag phase, followed
by exponential consumption, and a levelling off when the formose processes
runs out of formaldehyde supply. This is the point where the clear reaction
mixture starts to turn yellow and the generated \ch{C_4}--\ch{C_6} sugars
isomerize to a combinatorially complex mixture of compounds and black tar.
The core autocatalytic cycle of the formose process usually found in the
literature \cite{Ricardo:2006,Delidovich:2014,Ruiz-Mirazo:2014},
glycolaldehyde ``fixates'', via a series of keto-enol tautomerizations and
aldol additions, two formaldehyde molecules (thereby doubling in size)
followed by a retro-aldol fragmentation, resulting in two copies of
glycolaldehyde. However, experimental evidence exists
\cite{Ricardo:2006,Kim:2011} that this base cycle cannot account for the
massive consumption of formaldehyde, after the lag-phase. The reasons is
that, under the reaction conditions, enolization of the carbonyl group and
aldol addition are much faster than ``ketonization'' (restoring the
carbonyl group) required to close the autocatalytic base cycle. From this
experimental evidence the following mechanistic picture arises; fast
repeated addition of formaldehyde to enolized keto groups produces larger
sugars, draining material from the base cycle and retro-aldol fragmentation
of larger sugars replenishes the base cycle (or variants) with short
carbohydrates.  It is unknown how these higher order cycles are
structurally organized around the base autocatalytic cycle and how
they are interconnected.

For the formose chemistry we adopt the following naming scheme for
molecules; C$_{N\langle t\rangle}$, where $N$ specifies the number of
carbon atoms and $\langle t\rangle$ indicates the position of the double
bond.  We use $_\text{a}$ for aldehydes, $_\text{e}$ for enol forms, and
$_\text{k}$ for ketones. The model follows \cite{Benner:2010} using only
two basic types of reactions, keto-enol tautomerism and aldol reaction.
Both are reversible, thus giving the four transformation rules listed in
Tab.~\ref{tab:formoseRules} and the appendix. Starting
molecules are formaldehyde (\ch{C_1}) and glycolaldehyde (\ch{C_{2a}}).
The network is expanded to include all derivable molecules with at most 9
carbon atoms, thus reaching a size of 284 molecules and 978 reactions.  The
computation time was 8 seconds on a \computerDesc.
The ILP for finding pathways contains 33241 variables, of which 28240 are due to transit edges.
In this network all reactions are reversible and no transit node merging can thus been performed.

\begin{table}
\centering
\footnotesize
\begin{tabular}{@{}lll@{}}
\toprule
Abbr.	& Name	& Description \\
\midrule
KeEn	& Keto-Enol			& Keto to enol form conversion.
							\\
EnKe	& Enol-Keto			& Enol to keto form conversion.
							\\
AA		& Aldol addition		& Merge an enol and a keto.	
							\\
RAA	& Retro-aldol add.	& Split a keto into enol and keto.
							\\
\bottomrule
\end{tabular}
\caption{List of generic transformation rules for modelling the formose chemistry.}
\label{tab:formoseRules}
\end{table}

In the network we have enumerated overall autocatalytic pathways starting
from those of minimum size.  Specifically, pathways with the overall
reaction
\begin{align*}
\ch{C_{2a} + 2\ C_1 \carrow 2\ C_{2a}}
\end{align*}
with minimum number of unique reactions used.  For the purpose of this
enumeration we consider two solutions equivalent if the set of reactions
used is the same, thus how many times a reaction is used is ignored.
\begin{table}
\centering
\footnotesize
\begin{tabular}{@{}l@{\hspace{1em}}*{6}{r}@{}r@{}}
\toprule
	& \multicolumn{6}{c}{Maximum\ \#C}	\\
\cmidrule{2-7}
Unique reactions used & 4 & 5 & 6 & 7 & 8 & 9 & \multicolumn{1}{l}{\hspace{0.5em}Sum}	\\
\midrule
6       & 0     & 0     & 1     & 1     & 1     & 2     & 5     \\
7       & 0     & 0     & 0     & 0     & 0     & 2     & 2     \\
8       & 1     & 5     & 7     & 17    & 37    & 68    & 135   \\
9       & 0     & 0     & 12    & 12    & 37    & 69    & 130   \\
10      & 0     & 12    & 50    & 274   & 849   & ---     & $\geq 1185$  \\
11      & 0     & 5     & 41    & 190   & 738   & ---     & $\geq 974$   \\
\bottomrule
\end{tabular}
\caption[]{
  Overview of the number of autocatalytic flows in the formose chemistry.
  Solutions are grouped by the number of unique reactions used, and by the 
  number of carbon atoms in the largest molecule used.
  We were not able to compute the missing entries due the demand of 
  computation time (more than 200 hours) and memory (more than 64 GB RAM).
  Four of the smallest pathways can be found in the appendix.
}
\label{tab:formoseTable}
\end{table}
Tab.~\ref{tab:formoseTable} shows the resulting number of solutions found,
grouped by the number of reactions used and the maximum size of molecules
involved. The enumeration was split into 6 queries, one for each row of
the table, and the combined computation time was approximately 134 hours,
using a computing nodes with Intel\textregistered{}
Xeon\textsuperscript{TM} E5-2665 CPUs (\SI{2.4}{\giga\hertz}).  The
enumeration procedure issued in total 1,565,756 optimisation queries to
CPLEX, arising from a search trees of a combined size of 3,129,218
nodes. The unknown entries in the table are due to high memory demands
(more than \SI{64}{GB}) for single enumeration runs.
A selection of pathways can be found in the appendix.

The computational analysis of the chemical space of the formose process
reveals that the density of autocatalytic cycles is very high. The majority
of the enumerated autocatalytic cycles involve higher sugars
(\ch{C_5}--\ch{C_8}), but conform to the overall reaction of the shortest
possible overall autocatalytic cycle, referred to as base cycle. The
higher cycles branch off from compounds in the base cycle and merge back to
the base cycle further downstream. The resulting structure of interwoven
autocatalytic cycles is highly self-referential and shows, with respect to
this property, similarities to evolved metabolic networks, where also all
tentative waste compounds are recycled by feeding them back into the
metabolic network. The massive drain of material by the feeding of the
higher autocatalytic cycles considerably slows down the turn-over of the
base cycle or even result in breaking of the cycle. Furthermore, even if
the higher autocatalytic cycles themselves would not turn over, the base
cycle would still be replenished with \ch{C_2}/\ch{C_3} compounds generated
by retro-aldol fragmentation reactions of longer sugars. This results in a
mechanistic scenario where compounds of the base cycle massively fixate
formaldehyde in a fast polymerization type process to form longer sugars,
which themselves feedback to their starting points on lower levels via
fragmentation reactions, refilling these crucial compounds in the base
cycle. In that way the fragmentation reactions compensates for the material
loss of the autocatalytic base and higher cycles.

\section{Discussion}
The model presented here, based on integer hyperflows, provides a
versatile framework for querying reaction networks for pathways.  The
restriction to integers, although harder to solve in the general case,
has the advantage that the network flow solution can be directly
interpreted in terms of mechanisms.  Furthermore, questions such as
``how many of a specific product can be formed from a limited amount
of starting molecules'' can easily be formulated and answered in our
framework, due to the use of ILP.
This approach also enables a targeted enumeration of pathways of interest.
Naturally, such systematic enumerations can become
  computationally much more challenging than finding a single optimal
  solution. However, on the other hand, it allows for exploring the
  space of inferred mechanisms.

Autocatalysis, for instance, is frequently discussed as one of the key
mechanistic concepts to understand the transition from abiotic to biotic
chemistry. Reaction chemistries that ``maximize'' the emergence of this
reaction pattern are considered the most plausible predecessors of the
chemistries employed by present-day biochemistry.  To identify these
potential precursor reaction chemistries, a strict algorithmic approach for
the search and identification of functional subnetworks in arbitrary
reaction chemistry is indispensable. We applied our technique to the
formose process and found an intricate topological structure of cascading
autocatalytic cycles that feed upon each other, fitting well to the
existing experimental evidence.

The NOG problem is a prime example of the problem setting in engineering
cellular metabolisms, a branch of Synthetic Biology. Given starting
material, a target molecule, and a set of enzymes, the task is to find a
network that implements the desired chemical transformation. Because of
lack of efficient computational network design methods, the established
workflow rests on directed evolution and screening \cite{Boyle:2012}. This
requires searching for the desired transformation network in metabolic
networks of existing organisms, transplanting the identified pathways into
the microbial cell factory, followed by improving the performance of the
pathway in the alien environment of the host cell. Although impressive
examples of this approach have been published
\cite{Ro:2006,Thodey:2014,Fossati:2015}, this strategy must fail, if no
natural pathway is known that implements the desired transformation.

The pathway modelling framework presented here is implemented as part
  of a larger software package that includes the methods for automatic
  generation of reaction networks using graph transformation
  \cite{mod,mod0.5}.  As illustrated with both the formose and NOG
  chemistries, this combination results in an extremely powerful framework
for attacking a wide range of problems from Chemistry and Biology.  The
grounding of the graph transformation approach in well established
mathematical theory allows us to provide efficient algorithms for an
intermediary level of detail.  For instance, it gives us a handle on
tracking individual atoms throughout the network, e.g., following specific
pathways.  The graph transformation formalism also makes it possible to
lump reaction sequences into a single overall reaction
\cite{Andersen:2014c}, thereby enabling precise computer-assisted
coarse graining operations on reaction networks.

In recent years the interest in CTMs resurfaced in the context of designing
alternative networks performing the same function as their natural
archetypes \cite{Ebenhoeh:2003}, as well as in the context of the notion of
optimality in biochemical network structure. While the earliest work
focused on small, well-characterised pathways such as the pentose phosphate
pathway and the TCA cycle
\cite{Melendez-Hevia:1994},
recent work extends the original approaches to larger networks including
diverse reaction chemistry such as the central carbon metabolism
\cite{Noor:2010} or the \ch{CO_2} fixation pathways
\cite{Bar-Even:2010}. In this literature the concept of CTM is only
discussed implicitly and to our knowledge there has been no explicit
attempt to formalize CTMs.

Here, we have illustrated how higher-level CTMs can be detected efficiently
in given reaction networks using overall (auto)catalytic pathways as an
example. This is possible whenever input and output of a motif can be
specified as a linear constraint, which includes in particular sets of
allowed input molecules and desired output molecules. In combination with
the graph transformation formalism our framework is even capable of
designing alternative pathways enforcing, e.g., the use of prescribed
intermediates.
  
Last but not least, the use of directed multi-hypergraphs is sufficient to strictly enforce chemical constraints.
At the same time they make complex network transformations convenient to understand,
since hyperflows correspond to subnetworks that have intuitively understandable graphical representations.

\section*{Acknowledgments}
This work was supported in part by the Volkswagen Stiftung proj.~no.~I/82719,
the COST-Action CM1304 ``Systems Chemistry'', and
by the Danish Council for Independent Research, Natural
Sciences, grants DFF-1323-00247 and DFF-7014-00041.
It is also supported by the ELSI Origins Network (EON),
which is supported by a grant from the John Templeton Foundation.
The opinions expressed in this publication are those of the authors
and do not necessarily reflect the views of the John Templeton Foundation.

\bibliography{FlowILP}

\appendix

\section{Properties of the Expanded Hypergraph}
\label{sec:expandedProps}

This section is the full version, including proofs, of the corresponding
section in the paper.

\subsection{Size of the Expanded Network}
For a directed multi-hypergraph $\mathcal{H} = (V, E)$, let the size of it
be denoted by $\mathrm{size}(\mathcal{H}) = |V| + |E| + \sum_{e\in E}(|e^+|
+ |e^-|)$.  Note that if the hypergraph is seen as a bipartite normal
graph, then this size corresponds to the number of vertices and edges.

\begin{proposition}
  The size of the extended network and the expanded network is polynomial
  in the size of the original network.
\end{proposition}
\begin{proof}
  The size of the extended network is $size(\overline{\mathcal{H}}) =
  size(\mathcal{H}) + 4\cdot |V|$, as two half-edges are added to each
  vertex.  For the expanded network, $\widetilde{\mathcal{H}}$, the size
  depends on the in- and out-degree of the vertices in the extended
  network.  Let $d^-_{\overline E}(v)$ denote the in-degree of $v\in V$,
  and $d^+_{\overline E}(v)$ the out-degree.  Note that the degree counts
  the number of unique incident edges, so for $e\in \overline E, v\in V :
  m_v(e^-) > 1$ the size contribution of $e$ to $d^-_{\overline E}(v)$ is
  still only 1.  Then the size of the expanded network is
\begin{align*}
  size(\widetilde{\mathcal{H}})	&= size(\overline{\mathcal{H}}) - |V|		
  	\quad+
  \sum_{v\in V}\left(
  	d^-_{\overline E}(v) + d^+_{\overline E}(v)
  	+ 3\cdot
  	d^-_{\overline E}(v)\cdot d^+_{\overline E}(v)
	\right)
	\\
  &\leq size(\overline{\mathcal{H}}) - |V| + 2\cdot |V|\cdot |E| + 3\cdot
  |V|\cdot |E|^2
\end{align*}
where the inequality stems from the fact that at most all vertices are in
all head and tail sets, in the original network.
\end{proof}

\subsection{Translation of Flow}
\begin{proposition}
  A feasible flow $f\colon \widetilde{E}\rightarrow \mathbb{N}_0$ on
  $\widetilde{\mathcal{H}}$ can be converted into an equivalent feasible
  flow $g\colon \overline{E}\rightarrow \mathbb{N}_0$ in
  $\overline{\mathcal{H}}$, with: $g(e) = f(\widetilde{e})$, for all $e\in
  \overline{E}$.
\end{proposition}
\begin{proof}
  If $f$ is feasible, Eq.~\eqref{eq:masscons:expanded} holds for all
  $\widetilde{v}\in \widetilde{V}$. By the definition of
  $\widetilde{\mathcal{H}}$, we can say that Eq.~\eqref{eq:masscons:expanded}
  holds for all $\widetilde{v}\in V_v^-\cup V_v^+$ for all $v\in V$.
  Recall that all transit edges have singleton heads and tails, and $f(e) =
  f(\widetilde e), \forall e\in \overline E$.  Thus, by addition of
  Eq.~\eqref{eq:masscons:expanded} in each $v\in V$ we get that $\forall v\in V$:
\begin{align*}
0 &=
\sum_{u^-_{v,e}\in V_v^-}\left(
\overbrace{
	\sum_{u^+\in V^+_v}        f((u^-_{v,e}, u^+))
}^{\text{out-flow of } u^-_{v,e}}
-
\overbrace{
	m_{u^-_{v,e}}(e^-)f(e)
}^{\text{in-flow of } u^-_{v,e}}
\right)
\\
&+
\sum_{u^+_{v,e}\in V_v^+}\left(
\overbrace{
	m_{u^+_{v,e}}(e^+)f(e)
}^{\text{out-flow of } u^+_{v,e}}
-
\overbrace{
	\sum_{u^-\in V^-_v}		 f((u^-, u^+_{v,e}))
}^{\text{in-flow of } u^+_{v,e}}
\right) 				
\end{align*}
Here, the flow along each transit edge is first added and then subtracted
again, so we can simplify the expression to, $\forall v\in V$:
\begin{align*}
\sum_{u^-_{v,e}\in V_v^-} - m_{u^-_{v,e}}(e^-)f(e)
+
\sum_{u^+_{v,e}\in V_v^+} m_{u^+_{v,e}}(e^+)f(e)
= 0
\end{align*}
Using the definition of $V^-_v$ and $V^+_v$
(Eq.~\eqref{eq:expandedInVertices} and \eqref{eq:expandedOutVertices}) one
verifies that these relaxed constraints are exactly those of
Eq.~\eqref{eq:masscons}, i.e., the constraints on flows in
$\overline{\mathcal{H}}$.
\end{proof}

\begin{proposition}
  Let $f\colon \overline{E}\rightarrow \mathbb{N}_0$ be a feasible flow on
  $\overline{\mathcal{H}}$.  It can then be decided in polynomial time, in
  the size of $\mathcal{H}$, if a feasible flow $g\colon
  \widetilde{E}\rightarrow \mathbb{N}_0$ in $\widetilde{\mathcal{H}}$
  exists such that $g(\widetilde{e}) = f(e)$ for all $e\in \widetilde{E}$.
  If it exists it can be computed in polynomial time.
\end{proposition}
\begin{proof}
  The proof proceeds by a reduction to finding a feasible flow in bipartite
  normal directed graphs, with balance constraints. We refer to
  \cite{ahuja, digraphs} for a definition of this problem.  Recall that the
  edges of $\overline{H}$ are translated directly into a subset of the
  edges in $\widetilde{\mathcal{H}}$, and we as such are tasked with
  finding a feasible flow on all the transit edges, which can be decomposed
  into finding a feasible flow for each expanded vertex independently.  Let
  $v\in V$, then the hypergraph $(V^-_v\cup V^+_v, E_v)$ only contains
  edges with singleton head and tail multisets. It is therefore a normal
  directed, bipartite graph.  We then define the flow balance function
  $b\colon V^-_v\cup V^+_v\rightarrow \mathbb{N}_0$ as $\forall u_{v,
    e}^-\in V^-_v : b(u_{v, e}^-) = f(e)$ and $\forall u_{v, e}^+\in V^+_v
  : b(u_{v, e}^+) = -f(e)$.  Using the natural lower bound of flow $l\equiv
  0$ and infinite upper bound finally gives us the complete specification.  A
  feasible integer flow, if one exists, can be found in polynomial time in
  the size of the network\cite{ahuja, digraphs}.
\end{proof}

\section{Reduction from \textsc{Independent Set} to \textsc{Maximum Output}}
\label{sec:maxIndependent}
This reduction is intended to serve as an example for the section on comparison to existing methods, in the context of LP relaxation.
For a wider set of proofs of the computational complexity of integer hyperflow problems, see~\cite{andersen:12}, where also reductions to bounded hyperedge degree networks are described.
We here reduce between the optimisation versions of the problems, though it can easily be adopted to the decision versions.
In order to easily compare the reduction to the LP relaxation of the \textsc{Independent Set} problem, we will also state the ILP formulation of the problem.

\textsc{Independent Set}: Given an undirected graph $G$, find a set of vertices $V'\subseteq V(G)$, of maximum cardinality, such that no edge in the graph is between vertices of $V'$, i.e., $E(G)\cap V'\times V' = \emptyset$.

\textsc{Maximum Output}: Given 
\begin{itemize}
\item a directed multi-hypergraph $\mathcal{H}$,
\item a special output vertex $t\in V(\mathcal{H})$,
\item and lower and upper bounds on flow $l, u\colon \overline{E}(\mathcal{H})\rightarrow \mathbb{N}_0$,
\end{itemize}
find an integer hyperflow $f\colon \overline{E}(\mathcal{H})\rightarrow \mathbb{N}_0$, with maximum output of $t$, i.e.,  $\max f(e^+_t)$.

\subsection{ILP Formulation of \textsc{Independent Set}}
Let $G = (V, E)$ be the input graph.
\begin{align*}
	\max			&\quad \sum_{v\in V} x_v		\\
	\mathrm{s.t.}	&\quad x_u + x_v \leq 1 	&\forall (u, v)\in E 	\\
					&\quad x_v\in \{0, 1\} 	&\forall v\in V
\end{align*}
The resulting independent set is the vertices $v\in V$ with $x_v = 1$.

\subsection{Reduction to \textsc{Maximum Output}}
Let $G$ be the input graphs for the \textsc{Independent Set} problem.
We will first construct a directed multi-hypergraph $\mathcal{H}$:
\begin{align*}
	\mathcal{H} &= (V(\mathcal{H}), E(\mathcal{H})		\\
	V(\mathcal{H}) &= \{v_g\}\cup\{v_e\mid e\in E(G)\}	\\
	E(\mathcal{H} &= \{(\mset{v_e\mid e\in \delta(v)}, \mset{v_g}) \mid v\in V(H)\}
\end{align*}
We thus construct a vertex for each edge in $G$ and an extra ``goal vertex''.
The hyperedges correspond to the vertices of $G$, with the goal vertex as the head and the rest of the vertices corresponding to the incident edges of $G$ as the tail.

The hypergraph is then I/O-extended to $\mathcal{\overline{H}}$, and we define the lower bound on flow to be 0 on all hyperedges.
We set the upper bound to 0 for the input flow to the goal vertex, and 1 for the input to the remaining vertices. All other upper bounds are left infinite.

After maximisation of the output flow of the goal vertex we can construct the independent set as all the vertices $v\in V(G)$ where the corresponding hyperedge $e\in E(\mathcal{H})$ has $f(e) = 1$.

\section{Molecules}
\label{sec:molecules}
The following sections contains tables of the molecule abbreviations used
in the main text.  Some molecules are modelled using non-chemical vertex
labels that represent unimportant substructures.  For those molecules we
explicitly visualise the labelled graph, while for the strictly chemical
graphs we show a SMILES string.

\newcommand\gIn{\graphDFS[collapse hydrogens=false][scale=0.6, align=c, trim=0 -5pt 0 -5pt]}
\newcommand\litSmiles{\texttt}

\subsection{Molecule Abbreviations Rules for NOG}
\begin{adjustwidth}{-5cm}{-5cm}\scriptsize
\begin{center}
\begin{tabular}{@{}lll@{}}
\toprule
Abbreviations	& Name	& SMILES/Visualisation						\\
\midrule
AcCoA 			& Acetyl-CoA			& \gIn{CC(=O)S[CoA]} 				\\
AcP			& Acetyl phosphate	& \litSmiles{OP(O)(=O)OC(=O)C }	\\
\ch{C_8P}    	&						& \litSmiles{OCC(C(O)C(O)C(O)C(O)C(O)COP(O)(O)=O)=O}				\\
\ch{CO_2}		& Carbon dioxide		& \litSmiles{O=C=O}				\\
CoA			& Coenzyme A 			& \gIn{[CoA]S}						\\
DHAP			& Dihydroxyacetone phosphate	& \litSmiles{OP(O)(=O)OCC(=O)CO}						\\
E4P			& Erythrose 4-phosphate	& \litSmiles{OP(O)(=O)OCC(O)C(O)C=O}							\\
F6P			& Fructose 6-phosphate	& \litSmiles{OCC(=O)C(O)C(O)C(O)COP(=O)(O)O}					\\
FBP			& Fructose 1,6-bisphosphate	& \litSmiles{OC(COP(O)(O)=O)C(O)C(O)C(COP(O)(O)=O)=O}	\\
G3P			& Glyceraldehyde 3-phosphate	& \litSmiles{C(C(C=O)O)OP(=O)(O)O}						\\
\ch{H_2O}		& Water				& \litSmiles{O}						\\
\ch{P_i}		& Phosphate			& \litSmiles{O=P(O)(O)O}			\\
R5P			& Ribose 5-phosphate	& \litSmiles{OP(O)(=O)OCC(O)C(O)C(O)C=O}							\\
Ru5P, X5P 		& Ribulose 5-phosphate,
				Xylulose 5-phosphate	& \litSmiles{OCC(=O)C(O)C(O)COP(=O)(O)O}							\\
S7P			& Sedoheptulose 7-phosphate		& \litSmiles{O=P(O)(OCC(O)C(O)C(O)C(O)C(=O)CO)O}					\\
SBP			& Sedoheptulose 1,7-bisphosphate	& \litSmiles{OC(COP(O)(O)=O)C(O)C(O)C(O)C(COP(O)(O)=O)=O}	\\
\bottomrule
\end{tabular}
\end{center}
\end{adjustwidth}

\subsection*{Molecule Abbreviations for Formose}
\begin{center}\scriptsize
\begin{tabular}{@{}lll@{}}
\toprule
Abbreviations	& Name	& SMILES									\\
\midrule
\ch{C_1}		& Formaldehyde	& \litSmiles{C=O}									\\
\ch{C_{2a}}		& Glycolaldehyde	& \litSmiles{OCC=O}								\\
\ch{C_{2e}}		&					& \litSmiles{OC=CO}								\\
\ch{C_{3a}}		&					& \litSmiles{OCC(O)C=O}							\\
\bottomrule
\end{tabular}
\end{center}

\section{Transformation Rules}
\label{sec:rules}
The following sections contains visualisations of all graph transformation rules used to generated the analysed networks.
Each rule is annotated with references to the data its modelling is based on.
Most of these reference are to entries in the MACiE (Mechanism, Annotation, and Classification in Enzymes) database \cite{Holliday:2005,Holliday:2012}.

\newcommand\insertRule[2]{%
\subsubsection*{#2}%
\begin{center}%
	\small\ruleGML[]{rules/#1.gml}{\dpoRule[scale=0.6]}%
\end{center}%
}

\subsection*{Transformation Rules for NOG}
\insertRule{AL}{Aldolase}
MACiE entry 0052.

\insertRule{AlKe}{Aldose-Ketose}
MACiE entry 0308.

\insertRule{KeAl}{Ketose-Aldose}
MACiE entry 0308.

\insertRule{PHL}{Phosphohydrolase}
MACiE entries 0043, 0044, and 0047.

\insertRule{PK}{Phosphoketolase}
MetaCyc \cite{Caspi:2012} reaction entry 4.1.2.22.

\insertRule{TAL}{Transaldolase}
MACiE entry 0148.

\insertRule{TKL}{Transketolase}
MACiE entry 0219.

\subsection*{Transformation Rules for Formose}
The two reversible reactions have been modelled based on \cite{Benner:2010}.

\insertRule{KeEn}{Keto-Enol}
\insertRule{EnKe}{Enol-Keto}
\insertRule{AA}{Aldol addition}
\insertRule{RAA}{Retro-aldol addition}

\subsection*{Example Pathways in Formose}
In the following 4 pathways each molecule has a label on the form C$_{N\langle t\rangle}$,
where $N$ specifies the number of carbon atoms in the molecule and $\langle t\rangle$ indicates the position of the double bond.
We use $_\text{a}$ for aldehydes, $_\text{e}$ for enol forms, and $_\text{k}$ for ketones.
Therefore, formaldehyde is labelled with \ch{C_1} and glycolaldehyde with \ch{C_{2a}}.
Each reaction is labelled with the transformation rule used to generate the reaction (see previous section),
and with the flow through the reaction.

\newcommand\insertFormoseFig[3]{%
	\renewcommand\modInputPrefix{formose}%
	\renewcommand\modDGHyperImageScale{0.4}%
	\renewcommand\modDGHyperScale{1.10}%
	\subsubsection*{#2}
	\begin{adjustwidth}{-5cm}{-5cm}
	\begin{center}%
		\scriptsize%
		\input{#1}%
	\end{center}%
	\end{adjustwidth}%
	#3%
}

	\renewcommand\modInputPrefix{formose}%
	\renewcommand\modDGHyperImageScale{0.4}%
	\renewcommand\modDGHyperScale{1.10}%
	\subsubsection*{6 unique reactions, at most 6 carbons per molecule}
	\begin{adjustwidth}{-5cm}{-5cm}
	\begin{center}%
		\scriptsize%
		\input{formose/out/010_dg_0_1110_f_0_4_filt}	\end{center}%
	\end{adjustwidth}%
	This pathway is the shortest solution with fewest carbon atoms per molecule.%

	\renewcommand\modInputPrefix{formose}%
	\renewcommand\modDGHyperImageScale{0.4}%
	\renewcommand\modDGHyperScale{1.10}%
	\subsubsection*{7 unique reactions, at most 9 carbons per molecule, solution 1}
	\begin{adjustwidth}{-5cm}{-5cm}
	\begin{center}%
		\scriptsize%
		\input{formose/out/017_dg_0_1110_f_1_0_filt}%
	\end{center}%
	\end{adjustwidth}%
	\renewcommand\modInputPrefix{formose}%
	\renewcommand\modDGHyperImageScale{0.4}%
	\renewcommand\modDGHyperScale{1.10}%
	\subsubsection*{7 unique reactions, at most 9 carbons per molecule, solution 2}
	\begin{adjustwidth}{-5cm}{-5cm}
	\begin{center}%
		\scriptsize%
		\input{formose/out/026_dg_0_1110_f_1_1_filt}%
	\end{center}%
	\end{adjustwidth}%
	\renewcommand\modInputPrefix{formose}%
	\renewcommand\modDGHyperImageScale{0.4}%
	\renewcommand\modDGHyperScale{1.10}%
	\subsubsection*{8 unique reactions, at most 4 carbons per molecule}
	\begin{adjustwidth}{-5cm}{-5cm}
	\begin{center}%
		\scriptsize%
		\input{formose/out/029_dg_0_1110_f_2_4_filt}%
	\end{center}%
	\end{adjustwidth}%
	This is the pathway often depicted in the litterature.%

\end{document}

%% file: nogTable.tex
\setlength{\tabcolsep}{1.2ex}%
\begin{tabular}{@{}l@{\hspace{1em}}c@{}*{4}{c}@{}c@{}c@{}c@{}*{4}{c}@{}c@{}c@{}c@{}*{4}{c}@{}c@{}}
\toprule
     & \multicolumn{6}{@{}c@{}}{Only FBP} & & \multicolumn{13}{@{}c@{}}{Other Bisphosphates} \\
     \cmidrule{2-7}\cmidrule{9-21}
     & \multicolumn{6}{@{}c@{}}{8 Unique React.} & & \multicolumn{6}{@{}c@{}}{7 Unique React.} & & \multicolumn{6}{@{}c@{}}{8 Unique React.} \\
     \cmidrule{2-7}\cmidrule{9-14}\cmidrule{16-21}
PK Type     & \multicolumn{6}{@{}c@{}}{Reactions} & & \multicolumn{6}{@{}c@{}}{Reactions} & & \multicolumn{6}{@{}c@{}}{Reactions} \\
X, F, S, O      & \phantom{}& 8	& 9	& 10	& 11	 & \phantom{a} & \phantom{aa} & \phantom{}& 7	& 8	& 9	& 10	 & \phantom{a} & \phantom{aa} & \phantom{}& 8	& 9	& 10	& 11	 & \phantom{a}\\
\midrule
0, 0, 0, 3     & &  -- &  -- &  -- &  -- & & & & \cellcolor{lightgray!25}-- & \cellcolor{lightgray!25}-- & \cellcolor{lightgray!25}-- & \cellcolor{lightgray!25}-- & & & &  -- &  -- &   4 &  16 & \\
0, 0, 1, 2     & &  -- &  -- &  -- &  -- & & & & \cellcolor{lightgray!25}-- & \cellcolor{lightgray!25}-- & \cellcolor{lightgray!25}-- & \cellcolor{lightgray!25}-- & & & &  -- &   3 &   2 &  -- & \\
0, 0, 2, 1     & &  -- &  -- &  -- &  -- & & & & \cellcolor{lightgray!25}-- & \cellcolor{lightgray!25}-- & \cellcolor{lightgray!25}-- & \cellcolor{lightgray!25}-- & & & &  -- &   4 &  -- &  -- & \\
0, 0, 3, 0     & &  -- &  -- &   1 &   2 & & & & \cellcolor{lightgray!25}-- & \cellcolor{lightgray!25}-- & \cellcolor{lightgray!25}1 & \cellcolor{lightgray!25}2 & & & &  -- &  -- &   9 &  20 & \\
0, 1, 0, 2     & &  -- &  -- &  -- &  -- & & & & \cellcolor{lightgray!25}-- & \cellcolor{lightgray!25}-- & \cellcolor{lightgray!25}-- & \cellcolor{lightgray!25}-- & & & &  -- &   4 &   4 &  -- & \\
0, 1, 1, 1     & &  -- &  -- &  -- &  -- & & & & \cellcolor{lightgray!25}-- & \cellcolor{lightgray!25}-- & \cellcolor{lightgray!25}-- & \cellcolor{lightgray!25}-- & & & &   3 &  -- &  -- &  -- & \\
0, 1, 2, 0     & &  -- &   1 &  -- &  -- & & & & \cellcolor{lightgray!25}-- & \cellcolor{lightgray!25}1 & \cellcolor{lightgray!25}-- & \cellcolor{lightgray!25}-- & & & &  -- &   8 &   2 &  -- & \\
0, 2, 0, 1     & &  -- &  -- &  -- &  -- & & & & \cellcolor{lightgray!25}-- & \cellcolor{lightgray!25}-- & \cellcolor{lightgray!25}-- & \cellcolor{lightgray!25}-- & & & &  -- &   6 &  -- &  -- & \\
0, 2, 1, 0     & &  -- &   1 &  -- &  -- & & & & \cellcolor{lightgray!25}-- & \cellcolor{lightgray!25}1 & \cellcolor{lightgray!25}-- & \cellcolor{lightgray!25}-- & & & &  -- &   9 &  -- &  -- & \\
0, 3, 0, 0     & &  -- &  -- & \tikz[baseline=(A.base)]\node[draw, thick, text depth=0, inner sep=0, outer sep=0, minimum size=1.3em, fill=blue!20](A){2}; & 4$_{\text{a}}$ & & & & \cellcolor{lightgray!25}-- & \cellcolor{lightgray!25}-- & \tikz[baseline=(A.base)]\node[, text depth=0, inner sep=0, outer sep=0, minimum size=1.3em, fill=blue!20](A){\cellcolor{lightgray!25}2}; & \cellcolor{lightgray!25}4 & & & &  -- &  -- &  14 &  24 & \\
1, 0, 0, 2     & &  -- &  -- &  -- &  -- & & & & \cellcolor{lightgray!25}-- & \cellcolor{lightgray!25}-- & \cellcolor{lightgray!25}-- & \cellcolor{lightgray!25}-- & & & &  -- &   2 &   4 &  -- & \\
1, 0, 1, 1     & &  -- &  -- &  -- &  -- & & & & \cellcolor{lightgray!25}-- & \cellcolor{lightgray!25}-- & \cellcolor{lightgray!25}-- & \cellcolor{lightgray!25}-- & & & &   1 &  -- &  -- &  -- & \\
1, 0, 2, 0     & &  -- &   1 &  -- &  -- & & & & \cellcolor{lightgray!25}-- & \cellcolor{lightgray!25}1 & \cellcolor{lightgray!25}-- & \cellcolor{lightgray!25}-- & & & &  -- &   6 &   2 &  -- & \\
1, 1, 0, 1     & &  -- &  -- &  -- &  -- & & & & \cellcolor{lightgray!25}-- & \cellcolor{lightgray!25}-- & \cellcolor{lightgray!25}-- & \cellcolor{lightgray!25}-- & & & &   2 &  -- &  -- &  -- & \\
1, 1, 1, 0     & & \tikz[baseline=(A.base)]\node[, text depth=0, inner sep=0, outer sep=0, minimum size=1.3em, fill=green!20](A){1}; &  -- &  -- &  -- & & & & \tikz[baseline=(A.base)]\node[draw, thick, text depth=0, inner sep=0, outer sep=0, minimum size=1.3em, fill=green!20](A){\cellcolor{lightgray!25}1}; & \cellcolor{lightgray!25}-- & \cellcolor{lightgray!25}-- & \cellcolor{lightgray!25}-- & & & &   3 &  -- &  -- &  -- & \\
1, 2, 0, 0     & &  -- &   2 &  -- &  -- & & & & \cellcolor{lightgray!25}-- & \cellcolor{lightgray!25}2 & \cellcolor{lightgray!25}-- & \cellcolor{lightgray!25}-- & & & &  -- &  10 &  -- &  -- & \\
2, 0, 0, 1     & &  -- &  -- &  -- &  -- & & & & \cellcolor{lightgray!25}-- & \cellcolor{lightgray!25}-- & \cellcolor{lightgray!25}-- & \cellcolor{lightgray!25}-- & & & &  -- &   4 &  -- &  -- & \\
2, 0, 1, 0     & &  -- &   1 &  -- &  -- & & & & \cellcolor{lightgray!25}-- & \cellcolor{lightgray!25}1 & \cellcolor{lightgray!25}-- & \cellcolor{lightgray!25}-- & & & &  -- &   7 &  -- &  -- & \\
2, 1, 0, 0     & &  -- & 2$_{\text{c}}$ &  -- &  -- & & & & \cellcolor{lightgray!25}-- & \cellcolor{lightgray!25}2 & \cellcolor{lightgray!25}-- & \cellcolor{lightgray!25}-- & & & &  -- &  10 &  -- &  -- & \\
3, 0, 0, 0     & &  -- &  -- & 2$_{\text{b}}$ &   4 & & & & \cellcolor{lightgray!25}-- & \cellcolor{lightgray!25}-- & \cellcolor{lightgray!25}2 & \cellcolor{lightgray!25}4 & & & &  -- &  -- &  12 &  20 & \\
\bottomrule
\end{tabular}%

%% file: formose/out/010_dg_0_1110_f_0_4_filt.tex
\begin{tikzpicture}[scale=\modDGHyperScale]
\node[modStyleDGHyperVertex] (v-9-0) at (-7,-7) {\includegraphics[scale=\modDGHyperImageScale] {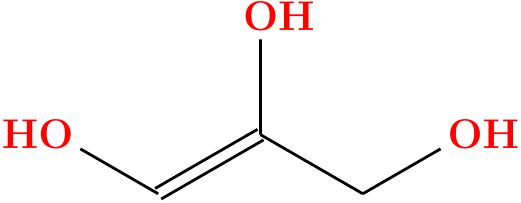}\\{$\mathrm{C_{3e}}$}};
\node[modStyleDGHyperVertex, above=of v-9-0] (v-0-0) {\includegraphics[scale=\modDGHyperImageScale] {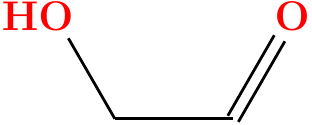}\\{$\mathrm{C_{2a}}$}};
\node[modStyleDGHyperEdge, above=of v-0-0] (v-282-0) {$\mathrm{RAA,\ 1}$};
\node[modStyleDGHyperVertex, above=11em of v-282-0] (v-1-0) {\includegraphics[scale=\modDGHyperImageScale] {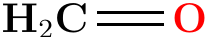}\\{$\mathrm{C_{1a}}$}};

\node[modStyleDGHyperEdge, left=2em of v-9-0] (v-277-0) {$\mathrm{RAA,\ 1}$};
\node[modStyleDGHyperVertex, left=2em of v-277-0)] (v-64-0) {\includegraphics[scale=\modDGHyperImageScale] {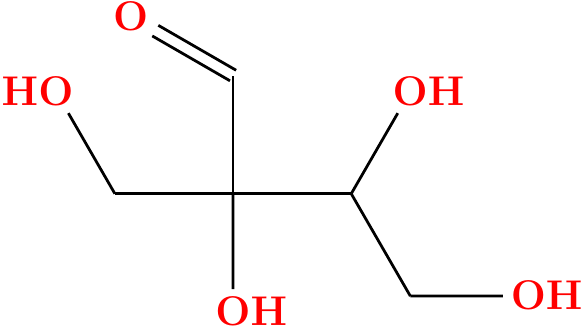}\\{$\mathrm{C_{5a}}$}};

\node[modStyleDGHyperEdge, right=2em of v-9-0] (v-70-0) {$\mathrm{AA,\ 1}$};
\node[modStyleDGHyperVertex, right=2em of v-70-0] (v-69-0) {\includegraphics[scale=\modDGHyperImageScale] {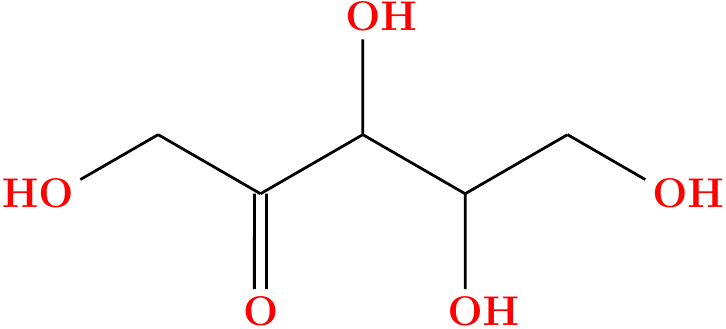}\\{$\mathrm{C_{5k}}$}};
\node[modStyleDGHyperVertex, above=of v-69-0] (v-19-0) {\includegraphics[scale=\modDGHyperImageScale] {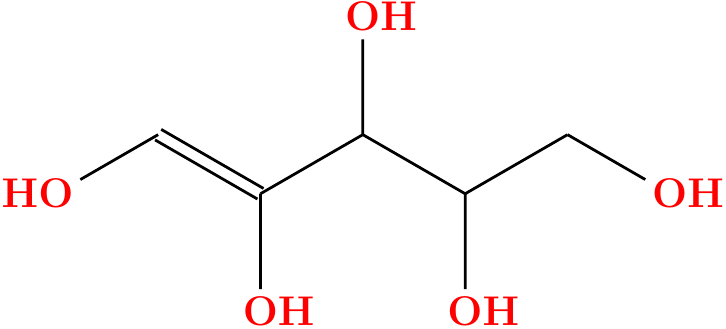}\\{$\mathrm{C_{5e}}$}};
\node[modStyleDGHyperEdge, above=of v-19-0] (v-240-0) {$\mathrm{AA,\ 1}$};
\node[modStyleDGHyperVertex, left=of v-240-0] (v-73-0) {\includegraphics[scale=\modDGHyperImageScale] {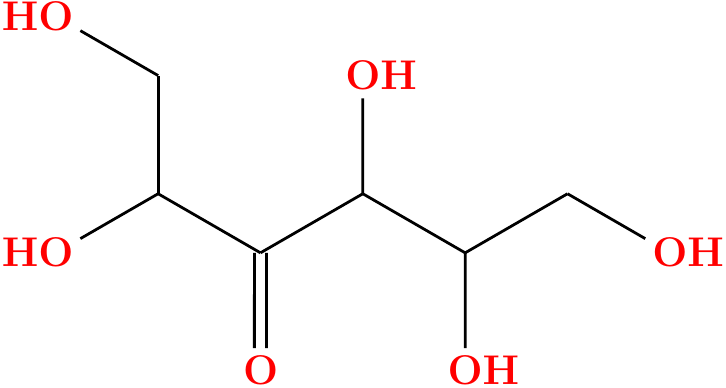}\\{$\mathrm{C_{6k}}$}};

\node[modStyleDGHyperEdge, at=(v-64-0 |- v-240-0)] (v-65-0) {$\mathrm{AA,\ 1}$};
\node[modStyleDGHyperVertex, right=of v-65-0] (v-11-0) {\includegraphics[scale=\modDGHyperImageScale] {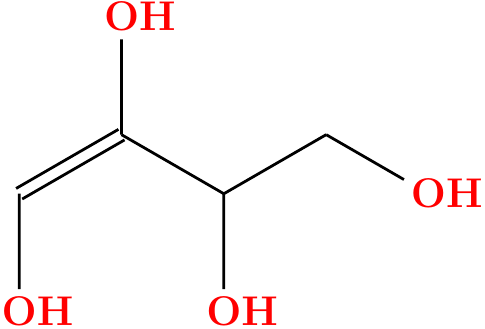}\\{$\mathrm{C_{4e}}$}};

\node[modStyleDGHyperVertexHiddenLarge, left=of v-0-0] (v-0-0-IOFlow) {};
\path[modStyleDGHyperConnector] (v-0-0-IOFlow) to[modStyleDGHyperHasReverseShortcut] node[auto, swap] {$\mathrm{1}$} (v-0-0);
\path[modStyleDGHyperConnector] (v-0-0) to[modStyleDGHyperHasReverseShortcut] node[auto, swap] {$\mathrm{2}$} (v-0-0-IOFlow);
\node[modStyleDGHyperVertexHiddenLarge, above=of v-1-0, overlay] (v-1-0-IOFlow) {};
\path[modStyleDGHyperConnector] (v-1-0-IOFlow) to node[auto, swap] {$\mathrm{2}$} (v-1-0);

\path[modStyleDGHyperConnector] (v-1-0) to[out=-178, in=80, out looseness=1.5, in looseness=0.3] (v-65-0);
\path[modStyleDGHyperConnector] (v-11-0) to (v-65-0);
\path[modStyleDGHyperConnector] (v-65-0) to (v-64-0);
\path[modStyleDGHyperConnector] (v-1-0) to[out=-2, in=110, out looseness=1.5, in looseness=0.3] (v-240-0);
\path[modStyleDGHyperConnector] (v-19-0) to (v-240-0);
\path[modStyleDGHyperConnector] (v-240-0) to (v-73-0);
\path[modStyleDGHyperConnector] (v-0-0) to (v-70-0);
\path[modStyleDGHyperConnector] (v-9-0) to (v-70-0);
\path[modStyleDGHyperConnector] (v-70-0) to (v-69-0);
\path[modStyleDGHyperConnector] (v-69-0) to node[auto, swap] {$\mathrm{KeEn,\ 1}$} (v-19-0);
\path[modStyleDGHyperConnector] (v-64-0) to (v-277-0);
\path[modStyleDGHyperConnector] (v-277-0) to (v-0-0);
\path[modStyleDGHyperConnector] (v-277-0) to (v-9-0);
\path[modStyleDGHyperConnector] (v-73-0) to (v-282-0);
\path[modStyleDGHyperConnector] (v-282-0) to (v-0-0);
\path[modStyleDGHyperConnector] (v-282-0) to (v-11-0);
\end{tikzpicture}%

%% file: formose/out/017_dg_0_1110_f_1_0_filt.tex
\begin{tikzpicture}[scale=\modDGHyperScale]
\node[modStyleDGHyperVertex] (v-211-0) {\includegraphics[scale=\modDGHyperImageScale] {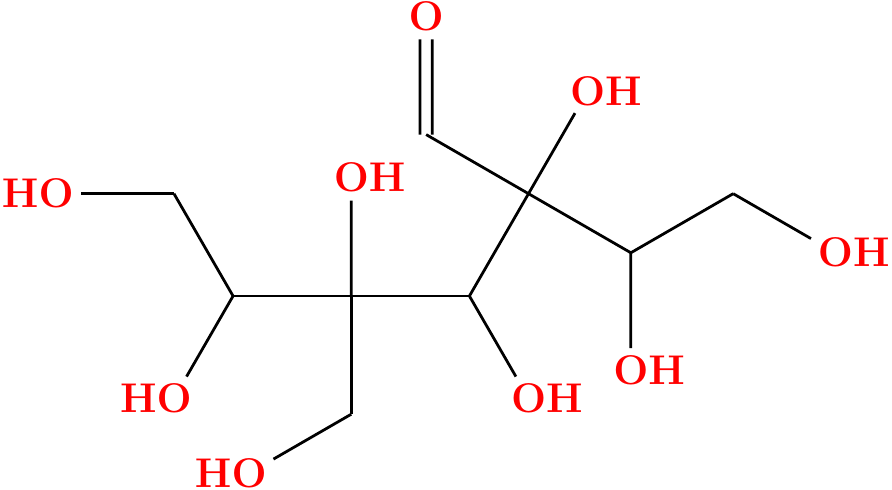}\\{$\mathrm{C_{9a}}$}};
\node[modStyleDGHyperEdge, left=2em of v-211-0] (v-956-0) {$\mathrm{AA,\ 1}$};
\node[modStyleDGHyperVertex, left=2em of v-956-0] (v-302-0) {\includegraphics[scale=\modDGHyperImageScale] {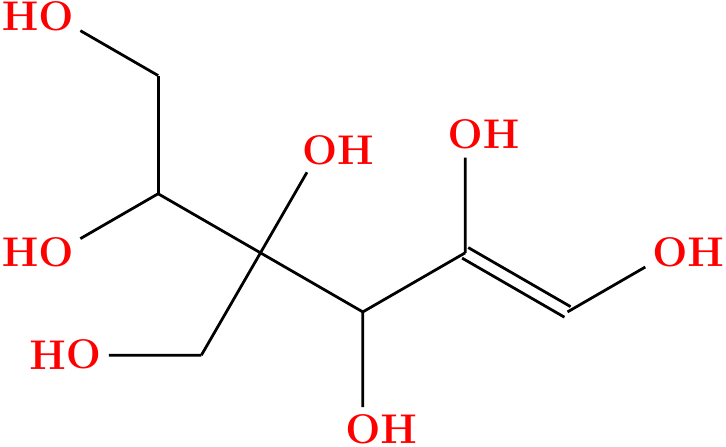}\\{$\mathrm{C_{7e}}$}};
\node[modStyleDGHyperVertex, above left=of v-302-0, xshift=2em, yshift=1em] (v-165-0) {\includegraphics[scale=\modDGHyperImageScale] {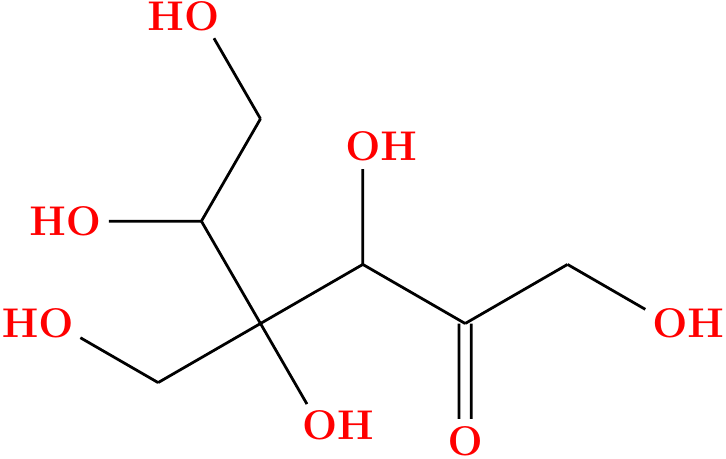}\\{$\mathrm{C_{7k}}$}};

\node[modStyleDGHyperEdge, above=2em of v-165-0] (v-166-0) {$\mathrm{AA,\ 1}$};
\node[modStyleDGHyperVertex, above=2em of v-166-0] (v-27-0) {\includegraphics[scale=\modDGHyperImageScale] {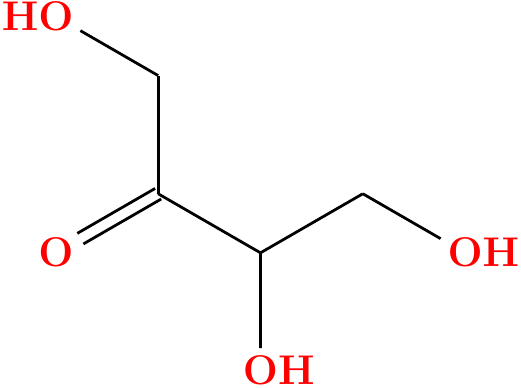}\\{$\mathrm{C_{4k}}$}};
\node[modStyleDGHyperEdge, right=2em of v-27-0] (v-63-0) {$\mathrm{AA,\ 1}$};

\node[modStyleDGHyperEdge, above=2em of v-211-0] (v-625-0) {$\mathrm{RAA,\ 1}$};
\node[modStyleDGHyperVertex, at=(v-27-0 -| v-625-0)] (v-11-0) {\includegraphics[scale=\modDGHyperImageScale] {\modInputPrefix/out/005_g_8_11111100.pdf}\\{$\mathrm{C_{4e}}$}};
\node[modStyleDGHyperEdge, left=2em of v-11-0] (v-65-0) {$\mathrm{AA,\ 1}$};

\node[modStyleDGHyperVertex, at=($(v-63-0)!0.5!(v-65-0)$), yshift=0em] (v-1-0) {\includegraphics[scale=\modDGHyperImageScale] {\modInputPrefix/out/003_g_0_11111100.pdf}\\{$\mathrm{C_{1a}}$}};

\node[modStyleDGHyperVertex, at=(v-65-0 |- v-625-0), yshift=8em] (v-64-0) {\includegraphics[scale=\modDGHyperImageScale] {\modInputPrefix/out/007_g_32_11111100.pdf}\\{$\mathrm{C_{5a}}$}};
\node[modStyleDGHyperVertex, right=2em of v-166-0] (v-9-0) {\includegraphics[scale=\modDGHyperImageScale] {\modInputPrefix/out/004_g_7_11111100.pdf}\\{$\mathrm{C_{3e}}$}};
\node[modStyleDGHyperEdge, at=($(v-64-0.west)!0.5!(v-9-0.east)$), yshift=-0em] (v-277-0) {$\mathrm{RAA,\ 2}$};

\node[modStyleDGHyperVertex, at=($(v-956-0)!0.5!(v-277-0)$)] (v-0-0) {\includegraphics[scale=\modDGHyperImageScale] {\modInputPrefix/out/002_g_1_11111100.pdf}\\{$\mathrm{C_{2a}}$}};

\node[modStyleDGHyperVertexHiddenLarge, left=of v-0-0, overlay] (v-0-0-IOFlow) {};
\path[modStyleDGHyperConnector] (v-0-0-IOFlow) to[modStyleDGHyperHasReverseShortcut] node[auto, swap] {$\mathrm{1}$} (v-0-0);
\path[modStyleDGHyperConnector] (v-0-0) to[modStyleDGHyperHasReverseShortcut] node[auto, swap] {$\mathrm{2}$} (v-0-0-IOFlow);
\node[modStyleDGHyperVertexHiddenLarge, above=of v-1-0, overlay] (v-1-0-IOFlow) {};
\path[modStyleDGHyperConnector] (v-1-0-IOFlow) to node[auto, swap] {$\mathrm{2}$} (v-1-0);

\path[modStyleDGHyperConnector] (v-1-0) to (v-63-0);
\path[modStyleDGHyperConnector] (v-9-0) to (v-63-0);
\path[modStyleDGHyperConnector] (v-63-0) to (v-27-0);
\path[modStyleDGHyperConnector] (v-1-0) to (v-65-0);
\path[modStyleDGHyperConnector] (v-11-0) to (v-65-0);
\path[modStyleDGHyperConnector] (v-65-0) to (v-64-0);
\path[modStyleDGHyperConnector] (v-9-0) to (v-166-0);
\path[modStyleDGHyperConnector] (v-27-0) to (v-166-0);
\path[modStyleDGHyperConnector] (v-166-0) to (v-165-0);
\path[modStyleDGHyperConnector] (v-64-0) to (v-277-0);
\path[modStyleDGHyperConnector] (v-277-0) to (v-0-0);
\path[modStyleDGHyperConnector] (v-277-0) to (v-9-0);
\path[modStyleDGHyperConnector] (v-165-0) to node[auto, swap] {$\mathrm{KeEn,\ 1}$} (v-302-0);
\path[modStyleDGHyperConnector] (v-211-0) to (v-625-0);
\path[modStyleDGHyperConnector] (v-625-0) to (v-11-0);
\path[modStyleDGHyperConnector] (v-625-0) to (v-64-0);
\path[modStyleDGHyperConnector] (v-0-0) to (v-956-0);
\path[modStyleDGHyperConnector] (v-302-0) to (v-956-0);
\path[modStyleDGHyperConnector] (v-956-0) to (v-211-0);
\end{tikzpicture}%

%% file: formose/out/026_dg_0_1110_f_1_1_filt.tex
\vspace{-4em}
\begin{tikzpicture}[scale=\modDGHyperScale]
\node[modStyleDGHyperVertex] (v-152-0) at (0, -10) {\includegraphics[scale=\modDGHyperImageScale] {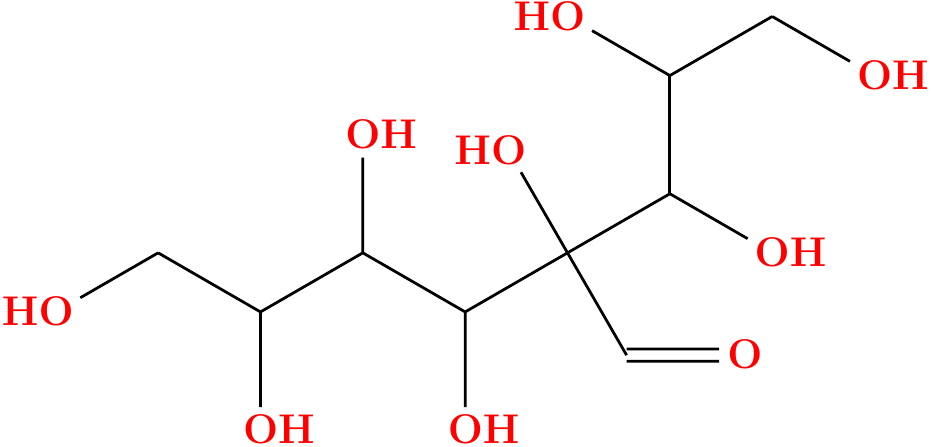}\\{$\mathrm{C_{9a}}$}};
\node[modStyleDGHyperEdge, left=4em of v-152-0] (v-153-0) {$\mathrm{AA,\ 1}$};
\node[modStyleDGHyperVertex, left=4em of v-153-0] (v-21-0) {\includegraphics[scale=\modDGHyperImageScale] {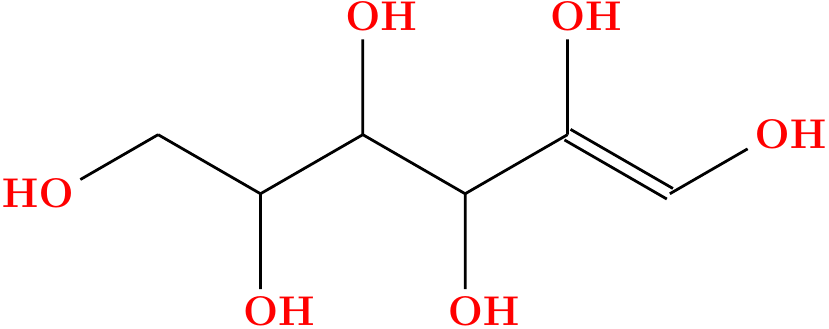}\\{$\mathrm{C_{6e}}$}};
\node[modStyleDGHyperVertex, above=5em of v-21-0] (v-150-0) {\includegraphics[scale=\modDGHyperImageScale] {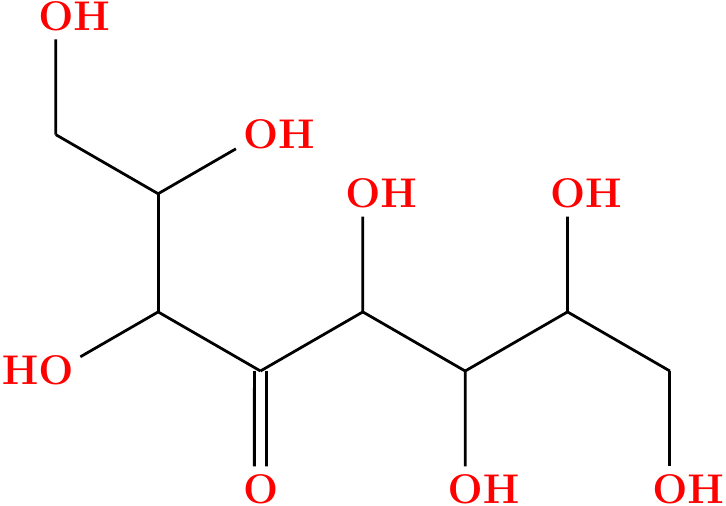}\\{$\mathrm{C_{8k}}$}};
\path (v-21-0) to
	node[modStyleDGHyperEdge, name=v-576-0] {$\mathrm{RAA,\ 1}$}
	(v-150-0);
\node[modStyleDGHyperVertex, above=of v-150-0, xshift=3em] (v-19-0) {\includegraphics[scale=\modDGHyperImageScale] {\modInputPrefix/out/006_g_11_11111100.pdf}\\{$\mathrm{C_{5e}}$}};

\node[modStyleDGHyperVertex, above=7em of v-153-0, xshift=8em] (v-5-0) {\includegraphics[scale=\modDGHyperImageScale] {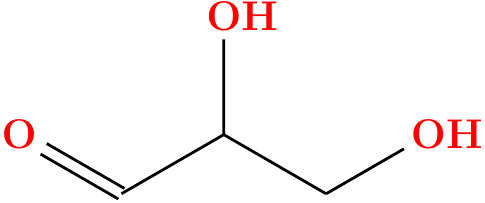}\\{$\mathrm{C_{3a}}$}};
\path (v-150-0) to
	node[modStyleDGHyperEdge, name=v-151-0] {$\mathrm{AA,\ 1}$}
	(v-5-0);

\node[modStyleDGHyperVertex, above=10em of v-5-0, xshift=-4em] (v-0-0) {\includegraphics[scale=\modDGHyperImageScale] {\modInputPrefix/out/002_g_1_11111100.pdf}\\{$\mathrm{C_{2a}}$}};
\node[modStyleDGHyperVertex, right= 5em of v-0-0] (v-2-0) {\includegraphics[scale=\modDGHyperImageScale] {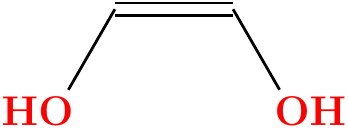}\\{$\mathrm{C_{2e}}$}};
\path (v-2-0) to
	node[modStyleDGHyperEdge, name=v-6-0] {$\mathrm{AA,\ 2}$}
	(v-5-0);
\node[modStyleDGHyperVertex, below right=of v-6-0] (v-1-0) {\includegraphics[scale=\modDGHyperImageScale] {\modInputPrefix/out/003_g_0_11111100.pdf}\\{$\mathrm{C_{1a}}$}};

\node[modStyleDGHyperEdge, at=($(v-0-0)!0.5!(v-2-0)$), yshift=4em] (v-18-0) {$\mathrm{RAA,\ 1}$};
\node[modStyleDGHyperVertex, above=2em of v-18-0] (v-7-0) {\includegraphics[scale=\modDGHyperImageScale] {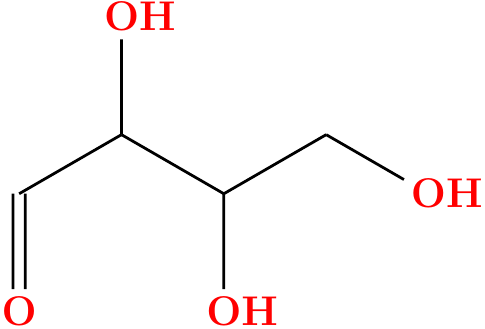}\\{$\mathrm{C_{4a}}$}};
\node[modStyleDGHyperEdge, right=of v-7-0] (v-578-0) {$\mathrm{RAA,\ 1}$};
\path[modStyleDGHyperConnector] (v-152-0) to[out=60, in=-70] (v-578-0);
\path[modStyleDGHyperConnector] (v-578-0) to (v-7-0);
\path[modStyleDGHyperConnector] (v-578-0) to[out=120, in=45] (v-19-0);
\path[modStyleDGHyperConnector] (v-150-0) to (v-576-0);
\path[modStyleDGHyperConnector] (v-576-0) to[out=5, in=-95, looseness=1.5] (v-0-0);
\path[modStyleDGHyperConnector] (v-576-0) to (v-21-0);

\node[modStyleDGHyperVertexHiddenLarge, left=of v-0-0, overlay] (v-0-0-IOFlow) {};
\path[modStyleDGHyperConnector] (v-0-0-IOFlow) to[modStyleDGHyperHasReverseShortcut] node[auto, swap] {$\mathrm{1}$} (v-0-0);
\path[modStyleDGHyperConnector] (v-0-0) to[modStyleDGHyperHasReverseShortcut] node[auto, swap] {$\mathrm{2}$} (v-0-0-IOFlow);
\node[modStyleDGHyperVertexHiddenLarge, below=of v-1-0, overlay] (v-1-0-IOFlow) {};
\path[modStyleDGHyperConnector] (v-1-0-IOFlow) to node[auto, swap] {$\mathrm{2}$} (v-1-0);

\path[modStyleDGHyperConnector] (v-0-0) to node[auto, swap] {$\mathrm{KeEn,\ 1}$} (v-2-0);
\path[modStyleDGHyperConnector] (v-1-0) to (v-6-0);
\path[modStyleDGHyperConnector] (v-2-0) to (v-6-0);
\path[modStyleDGHyperConnector] (v-6-0) to (v-5-0);
\path[modStyleDGHyperConnector] (v-7-0) to (v-18-0);
\path[modStyleDGHyperConnector] (v-18-0) to (v-0-0);
\path[modStyleDGHyperConnector] (v-18-0) to (v-2-0);
\path[modStyleDGHyperConnector] (v-5-0) to (v-151-0);
\path[modStyleDGHyperConnector] (v-19-0) to (v-151-0);
\path[modStyleDGHyperConnector] (v-151-0) to (v-150-0);
\path[modStyleDGHyperConnector] (v-5-0) to (v-153-0);
\path[modStyleDGHyperConnector] (v-21-0) to (v-153-0);
\path[modStyleDGHyperConnector] (v-153-0) to (v-152-0);
\end{tikzpicture}%

%% file: formose/out/029_dg_0_1110_f_2_4_filt.tex
\begin{tikzpicture}[scale=\modDGHyperScale]
\node[modStyleDGHyperVertex] (v-0-0) {\includegraphics[scale=\modDGHyperImageScale] {\modInputPrefix/out/002_g_1_11111100.pdf}\\{$\mathrm{C_{2a}}$}};
\node[modStyleDGHyperVertex, above=of v-0-0] (v-2-0) {\includegraphics[scale=\modDGHyperImageScale] {\modInputPrefix/out/020_g_4_11111100.pdf}\\{$\mathrm{C_{2e}}$}};
\node[modStyleDGHyperEdge, at=($(v-0-0)!0.5!(v-2-0)$), xshift=8em] (v-18-0) {$\mathrm{RAA,\ 1}$};
\node[modStyleDGHyperVertex, right=of v-18-0] (v-7-0) {\includegraphics[scale=\modDGHyperImageScale] {\modInputPrefix/out/022_g_6_11111100.pdf}\\{$\mathrm{C_{4a}}$}};
\node[modStyleDGHyperVertex, right=5em of v-7-0] (v-11-0) {\includegraphics[scale=\modDGHyperImageScale] {\modInputPrefix/out/005_g_8_11111100.pdf}\\{$\mathrm{C_{4e}}$}};
\node[modStyleDGHyperVertex, above=of v-11-0] (v-27-0) {\includegraphics[scale=\modDGHyperImageScale] {\modInputPrefix/out/013_g_14_11111100.pdf}\\{$\mathrm{C_{4k}}$}};
\node[modStyleDGHyperEdge, at=(v-2-0 |- v-27-0)] (v-6-0) {$\mathrm{AA,\ 1}$};
\node[modStyleDGHyperVertex, at=($(v-6-0)!0.33!(v-27-0)$)] (v-1-0) {\includegraphics[scale=\modDGHyperImageScale] {\modInputPrefix/out/003_g_0_11111100.pdf}\\{$\mathrm{C_{1a}}$}};
\node[modStyleDGHyperEdge, , at=($(v-6-0)!0.66!(v-27-0)$)] (v-63-0) {$\mathrm{AA,\ 1}$};

\node[modStyleDGHyperVertex, above=of v-6-0] (v-5-0) {\includegraphics[scale=\modDGHyperImageScale] {\modInputPrefix/out/021_g_5_11111100.pdf}\\{$\mathrm{C_{3a}}$}};
\node[modStyleDGHyperVertex, above=of v-63-0] (v-9-0) {\includegraphics[scale=\modDGHyperImageScale] {\modInputPrefix/out/004_g_7_11111100.pdf}\\{$\mathrm{C_{3e}}$}};

\node[modStyleDGHyperVertexHiddenLarge, below=of v-0-0, overlay] (v-0-0-IOFlow) {};
\path[modStyleDGHyperConnector] (v-0-0-IOFlow) to[modStyleDGHyperHasReverseShortcut] node[auto, swap] {$\mathrm{1}$} (v-0-0);
\path[modStyleDGHyperConnector] (v-0-0) to[modStyleDGHyperHasReverseShortcut] node[auto, swap] {$\mathrm{2}$} (v-0-0-IOFlow);
\node[modStyleDGHyperVertexHiddenLarge, below=of v-1-0] (v-1-0-IOFlow) {};
\path[modStyleDGHyperConnector] (v-1-0-IOFlow) to node[auto, swap] {$\mathrm{2}$} (v-1-0);

\path[modStyleDGHyperConnector] (v-0-0) to[modStyleDGHyperHasReverseShortcut] node[auto, swap] {$\mathrm{KeEn,\ 1}$} (v-2-0);
\path[modStyleDGHyperConnector] (v-2-0) to[modStyleDGHyperHasReverseShortcut] node[auto, swap] {$\mathrm{EnKe,\ 1}$} (v-0-0);
\path[modStyleDGHyperConnector] (v-1-0) to (v-6-0);
\path[modStyleDGHyperConnector] (v-2-0) to (v-6-0);
\path[modStyleDGHyperConnector] (v-6-0) to (v-5-0);
\path[modStyleDGHyperConnector] (v-5-0) to node[auto, swap] {$\mathrm{KeEn,\ 1}$} (v-9-0);
\path[modStyleDGHyperConnector] (v-7-0) to (v-18-0);
\path[modStyleDGHyperConnector] (v-18-0) to (v-0-0);
\path[modStyleDGHyperConnector] (v-18-0) to (v-2-0);
\path[modStyleDGHyperConnector] (v-11-0) to node[auto, swap] {$\mathrm{EnKe,\ 1}$} (v-7-0);
\path[modStyleDGHyperConnector] (v-1-0) to (v-63-0);
\path[modStyleDGHyperConnector] (v-9-0) to (v-63-0);
\path[modStyleDGHyperConnector] (v-63-0) to (v-27-0);
\path[modStyleDGHyperConnector] (v-27-0) to node[auto, swap] {$\mathrm{KeEn,\ 1}$} (v-11-0);
\end{tikzpicture}%

%% file: FlowILP.bbl
\begin{thebibliography}{10}

\bibitem{ahuja}
R.~K. Ahuja, T.~L. Magnanti, and J.B. Orlin.
\newblock {\em Network Flows: Theory, Algorithms, and Applications}.
\newblock Prentice Hall, Englewood Cliffs, NJ, 1993.

\bibitem{andersen:12}
J.~L. Andersen, C.~Flamm, D.~Merkle, and P.~F. Stadler.
\newblock Maximizing output and recognizing autocatalysis in chemical reaction
  networks is {NP}-complete.
\newblock {\em J. Systems Chem.}, 3:1, 2012.

\bibitem{mod}
Jakob~L. Andersen.
\newblock {Med\O lDatschgerl (M\O D)}.
\newblock \url{http://mod.imada.sdu.dk}, 2017.

\bibitem{hcn}
Jakob~L. Andersen, Tommy Andersen, Christoph Flamm, Martin~M. Hanczyc, Daniel
  Merkle, and Peter~F. Stadler.
\newblock Navigating the chemical space of hcn polymerization and hydrolysis:
  Guiding graph grammars by mass spectrometry data.
\newblock {\em Entropy}, 15(10):4066--4083, 2013.

\bibitem{dna}
Jakob~L. Andersen, Christoph Flamm, Martin~M. Hanczyc, and Daniel Merkle.
\newblock Towards an optimal {DNA}-templated molecular assembler.
\newblock In {\em ALIFE 14: The Fourteenth Conference on the Synthesis and
  Simulation of Living Systems}, volume~14, pages 557--564, 2014.

\bibitem{Andersen:13a}
Jakob~L Andersen, Christoph Flamm, Daniel Merkle, and Peter~F. Stadler.
\newblock Inferring chemical reaction patterns using graph grammar rule
  composition.
\newblock {\em J. Syst. Chem.}, 4:4, 2013.

\bibitem{Andersen:2014c}
Jakob~L Andersen, Christoph Flamm, Daniel Merkle, and Peter~F Stadler.
\newblock 50 shades of rule composition: {F}rom chemical reactions to higher
  levels of abstraction.
\newblock In Fran{\c{c}}ois Fages and Carla Carla~Piazza, editors, {\em
  Proceedings of the 1st International Conference on Formal Methods in
  Macro-Biology}, volume 8738 of {\em LNCS}, pages 117--135. Springer-Verlag,
  Berlin Heidelberg, 2014.

\bibitem{dgStrat}
Jakob~L. Andersen, Christoph Flamm, Daniel Merkle, and Peter~F. Stadler.
\newblock Generic strategies for chemical space exploration.
\newblock {\em International Journal of Computational Biology and Drug Design},
  7(2/3):225 -- 258, 2014.

\bibitem{eschenmoser}
Jakob~L. Andersen, Christoph Flamm, Daniel Merkle, and Peter~F. Stadler.
\newblock In silico support for {Eschenmoser{\textquoteright}s} glyoxylate
  scenario.
\newblock {\em Isr. J. Chem.}, 2015.

\bibitem{mod0.5}
Jakob~L. Andersen, Christoph Flamm, Daniel Merkle, and Peter~F. Stadler.
\newblock A software package for chemically inspired graph transformation.
\newblock In Rachid Echahed and Mark Minas, editors, {\em Graph Transformation:
  9th International Conference, ICGT 2016, in Memory of Hartmut Ehrig, Held as
  Part of STAF 2016, Vienna, Austria, July 5-6, 2016, Proceedings}, pages
  73--88, Cham, 2016. Springer International Publishing.

\bibitem{digraphs}
J{\o}rgen Bang-Jensen and Gregory~Z Gutin.
\newblock Digraphs: Theory, algorithms and applications.
\newblock {\em Springer Monographs in Mathematics}, 2009.

\bibitem{Bar-Even:2010}
Arren Bar-Even, Elad Noor, Nathan~E. Lewis, and Ron Milo.
\newblock Design and analysis of synthetic carbon fixation pathways.
\newblock {\em Proc Natl Acad Sci}, 107:8889--8894, 2010.

\bibitem{Beasley:01012007}
John~E. Beasley and Francisco~J. Planes.
\newblock Recovering metabolic pathways via optimization.
\newblock {\em Bioinformatics}, 23(1):92--98, 2007.

\bibitem{Benner:2010}
S.~A. Benner, H.-J. Kim, M.-J. Kim, and A.~Ricardo.
\newblock Planetary organic chemistry and the origins of biomolecules.
\newblock {\em Cold Spring Harbor Perspect. Biol.}, 2(7), 2010.

\bibitem{Bissette:2013}
Andrew~J. Bissette and Stephen~P. Fletcher.
\newblock Mechanisms of {A}utocatalysis.
\newblock {\em J Angew Chemie Int Ed}, 52(49):12800--12826, 2013.

\bibitem{blackmond:2009}
DG~Blackmond.
\newblock An examination of the role of autocatalytic cycles in the chemistry
  of proposed primordial reactions.
\newblock {\em J Angew Chemie Int Ed}, 48:386--390, 2009.

\bibitem{bogorad2013synthetic}
Igor~W Bogorad, Tzu-Shyang Lin, and James~C Liao.
\newblock Synthetic non-oxidative glycolysis enables complete carbon
  conservation.
\newblock {\em Nature}, 502(7473):693--697, 2013.

\bibitem{Boyle:2012}
Patrick~M. Boyle and Pamela~A. Silver.
\newblock Parts plus pipes: Synthetic biology approaches to metabolic
  engineering.
\newblock {\em Metab. Eng.}, 14:223--232, 2012.

\bibitem{Burgard2001}
Anthony~P. Burgard, Shankar Vaidyaraman, and Costas~D. Maranas.
\newblock Minimal reaction sets for escherichia coli metabolism under different
  growth requirements and uptake environments.
\newblock {\em Biotechnol. Progr.}, 17(5):791--797, 2001.

\bibitem{Butlerov:1861}
Alexandr~Mikhaylovich Butlerov.
\newblock Einiges {\"u}ber die chemische {S}tructur der {K}{\"o}rper.
\newblock {\em Zeitschrift f{\"u}r Chemie}, 4:549--560, 1861.

\bibitem{Cambini:97}
Riccardo Cambini, Giorgio Gallo, and Maria~Grazia Scutell{\`a}.
\newblock Flows on hypergraphs.
\newblock {\em Math. Program.}, 78:195--217, 1997.

\bibitem{Caspi:2012}
R~Caspi, T~Altman, K~Dreher, C~A Fulcher, P~Subhraveti, I~M Keseler, A~Kothari,
  M~Krummenacker, M~Latendresse, L~A Mueller, Q~Ong, S~Paley, A~Pujar, A~G
  Shearer, M~Travers, D~Weerasinghe, P~Zhang, and P~D Karp.
\newblock The \texttt{MetaCyc} database of metabolic pathways and enzymes and
  the \texttt{BioCyc} collection of pathway/genome databases.
\newblock {\em Nucleic Acids Res.}, 40:D742--D753, 2012.

\bibitem{Centler:08}
Florian Centler, Christoph Kaleta, Pietro Speroni~di Fenizio, and Peter
  Dittrich.
\newblock Computing chemical organizations in biological networks.
\newblock {\em Bioinformatics}, 24:1611--1618, 2008.

\bibitem{Choi:2015}
Jung-Min Choi, Sang-Soo Han, and Hak-Sung Kim.
\newblock Industrial applications of enzyme biocatalysis: {C}urrent status and
  future aspects.
\newblock {\em Biotech Adv}, 33:1443--1454, 2015.

\bibitem{Figueiredo:2009}
Luis~F. de~Figueiredo, Adam Podhorski, Angel Rubio, Christoph Kaleta, John~E.
  Beasley, Stefan Schuster, and Francisco~J. Planes.
\newblock Computing the shortest elementary flux modes in genome-scale
  metabolic networks.
\newblock {\em Bioinformatics}, 25(23):3158--3165, 2009.

\bibitem{Delidovich:2014}
Irina~V. Delidovich, Alexandr~N. Simonov, Oxana~P. Taran, and Valentin~N.
  Parmon.
\newblock Catalytic {F}ormation of {M}onosaccharides: {F}rom the {F}ormose
  {R}eaction towards {S}elective {S}ynthesis.
\newblock {\em ChemSusChem}, 7(7):1833--1846, 2014.

\bibitem{DeLoache:2015}
William~C DeLoache, Zachary~N Russ, Lauren Narcross, Andrew~M Gonzales, Vincent
  J~J Martin, and John~E Dueber.
\newblock An enzyme-coupled biosensor enables (s)-reticuline production in
  yeast from glucose.
\newblock {\em Nature Chem Biol}, 11:465--471, 2015.

\bibitem{Ebenhoeh:2003}
Oliver Ebenh{\"o}h and Reinhart Heinrich.
\newblock Stoichiometric design of metabolic networks: multifunctionality,
  clusters, optimization, weak and strong robustness.
\newblock {\em Bull. Math. Biol.}, 65:323--357, 2003.

\bibitem{Fedoroff:02}
Nina Fedoroff and Walter Fontana.
\newblock Small numbers of big molecules.
\newblock {\em Science}, 297:1129--1130, 2002.

\bibitem{Fell:86}
D~A Fell and J~R Small.
\newblock Fat synthesis in adipose tissue. an examination of stoichiometric
  constraints.
\newblock {\em Biochem. J.}, 238:781--786, 1986.

\bibitem{Fossati:2015}
Elena Fossati, Lauren Narcross, Andrew Ekins, Jean-Pierre Falgueyret, and
  Vincent J.~J. Martin.
\newblock Synthesis of morphinan alkaloids in saccharomyces cerevisiae.
\newblock {\em PLoS One}, 10(4):e0124459, 2015.

\bibitem{Gallo:98a}
G.~Gallo, C~Gentile, D~Pretolani, and G.~Rago.
\newblock Max {Horn SAT} and the minimum cut problem in directed hypergraphs.
\newblock {\em Math. Programming}, 80:213--237, 1998.

\bibitem{Gallo:93}
Giorgio Gallo, Giustino Longo, Stefano Pallottino, and Sang Nguyen.
\newblock Directed hypergraphs and applications.
\newblock {\em Discrete Appl. Math.}, 42(2–3):177--201, 1993.

\bibitem{Garcia-Junceda:2015}
Eduardo Garc{\'i}a-Junceda, Iv{\'a}n Lavandera, D{\"o}rte Rother, and Joerg~H
  Schrittwieser.
\newblock (chemo)enzymatic cascades -- nature's synthetic strategy transferred
  to the laboratory.
\newblock {\em J Mol Cat B: Enzymatic}, 114:1--6, 2015.

\bibitem{Gillespie:07}
D~T Gillespie.
\newblock Stochastic simulation of chemical kinetics.
\newblock {\em Annu. Rev. Phys. Chem.}, 58:35--55, 2007.

\bibitem{lownumbers1}
Purnananda Guptasarma.
\newblock Does replication-induced transcription regulate synthesis of the
  myriad low copy number proteins of escherichia coli?
\newblock {\em Bioessays}, 17(11):987--997, 1995.

\bibitem{scope1}
Thomas Handorf, Oliver Ebenhöh, and Reinhart Heinrich.
\newblock Expanding metabolic networks: Scopes of compounds, robustness, and
  evolution.
\newblock {\em J. Mol. Evol.}, 61:498--512, 2005.
\newblock 10.1007/s00239-005-0027-1.

\bibitem{Holliday:2012}
Gemma~L Holliday, Claudia Andreini, Julia~D Fischer, Syed~Asad Rahman, Daniel~E
  Almonacid, Sophie~T Williams, and William~R Pearson.
\newblock {MACiE}: exploring the diversity of biochemical reactions.
\newblock {\em Nucleic Acids Res.}, 40:D783--D789, 2012.

\bibitem{Holliday:2005}
Gemma~L. Holliday, Gail~J. Bartlett, Daniel~E. Almonacid, Noel~M. O'Boyle,
  Peter Murray-Rust, Janet~M. Thornton, and John B.~O. Mitchell.
\newblock {MACiE}: a database of enzyme reaction mechanisms.
\newblock {\em Bioinformatics}, 21:4315--4316, 2005.

\bibitem{Hordijk:13}
Wim Hordijk.
\newblock {A}utocatalytic {S}ets: {F}rom the {O}rigin of {L}ife to the
  {E}conomy.
\newblock {\em BioSci}, 63(11):877--881, 2013.

\bibitem{Jonnalagadda2014}
Sudhakar Jonnalagadda and Rajagopalan Srinivasan.
\newblock An efficient graph theory based method to identify every minimal
  reaction set in a metabolic network.
\newblock {\em BMC Syst. Biol.}, 8(1):28, Mar 2014.

\bibitem{Kaleta:06}
C~Kaleta, F~Centler, and P~Dittrich.
\newblock Analyzing molecular reaction networks: from pathways to chemical
  organizations.
\newblock {\em Mol. Biotechnol.}, 34:117--123, 2006.

\bibitem{Kaleta01102009}
Christoph Kaleta, Luís~Filipe de~Figueiredo, and Stefan Schuster.
\newblock Can the whole be less than the sum of its parts? pathway analysis in
  genome-scale metabolic networks using elementary flux patterns.
\newblock {\em Genome Res.}, 19(10):1872--1883, 2009.

\bibitem{lpPoly2}
Narendra Karmarkar.
\newblock A new polynomial-time algorithm for linear programming.
\newblock In {\em Proceedings of the sixteenth annual ACM symposium on Theory
  of computing}, pages 302--311. ACM, 1984.

\bibitem{Karp:72}
R.~M. Karp.
\newblock Reducibility among combinatorial problems.
\newblock In R.~E. Miller and J.~W. Thatcher, editors, {\em Complexity of
  Computer Computations}, pages 85--103, NY, 1972. Plenum Press.

\bibitem{Kauffman:03}
Kenneth~J Kauffman, Purusharth Prakash, and Jeremy~S Edwards.
\newblock Advances in flux balance analysis.
\newblock {\em Current Opinion Biotech.}, 14:491--496, 2003.

\bibitem{Kauffman:86}
S.~A. Kauffman.
\newblock Autocatalytic sets of proteins.
\newblock {\em J. Theor. Biol.}, 119(1):1 -- 24, 1986.

\bibitem{lpPoly1}
Leonid~G Khachiyan.
\newblock Polynomial algorithms in linear programming.
\newblock {\em USSR Computational Mathematics and Mathematical Physics},
  20(1):53--72, 1980.

\bibitem{Kim:2011}
Hyo-Joong Kim, Alonso Ricardo, Heshan~I. Illangkoon, Myong~Jung Kim, Matthew~A.
  Carrigan, Fabianne Frye, and Steven~A. Benner.
\newblock Synthesis of {C}arbohydrates in {M}ineral-{G}uided {P}rebiotic
  {C}ycles.
\newblock {\em J. Am. Chem. Soc.}, 133(24):9457--9468, 2011.

\bibitem{Klamt:02}
S.~Klamt and J.~Stelling.
\newblock Combinatorial complexity of pathway analysis in metabolic networks.
\newblock {\em Mol. Biol. Rep.}, 29:233--236, 2002.

\bibitem{Klamt:03}
Steffen Klamt and J{\"o}rg Stelling.
\newblock Two approaches for metabolic pathway analysis?
\newblock {\em Trends Biotechnol.}, 21:64--69, 2003.

\bibitem{Kleijn:2005}
Roelco~J Kleijn, Wouter~A van Winden, Walter~M van Gulik, and Joseph~J Heijnen.
\newblock Reviciting the ${}^{13}${C}-label disribution of the non-oxidative
  branch of pentose phosphate pathway based upon kinetic and genetic evidence.
\newblock {\em FEBS J.}, 272:4970--4982, 2005.

\bibitem{Kreyssig:14}
Peter Kreyssig, Christian Wozar, Stephan Peter, Tom{\'a}s Veloz, Bashar
  Ibrahim, and Peter Dittrich.
\newblock Effects of small particle numbers on long-term behaviour in discrete
  biochemical systems.
\newblock {\em Bioinformatics}, 30:i475--i481, 2014.

\bibitem{kun:08}
{\'A}d{\'a}m Kun, Bal{\'a}zs Papp, and E{\"o}rs Szathm{\'a}ry.
\newblock Computational identification of obligatorily autocatalytic
  replicators embedded in metabolic networks.
\newblock {\em Genome Biol.}, 9:R51, 2008.

\bibitem{fbaEnum}
Sangbum Lee, Chan Phalakornkule, Michael~M Domach, and Ignacio~E Grossmann.
\newblock Recursive milp model for finding all the alternate optima in lp
  models for metabolic networks.
\newblock {\em Computers \& Chemical Engineering}, 24(2):711--716, 2000.

\bibitem{Melendez-Hevia:1994}
Enrique Mel{\'e}ndez-Hevia, Thomas~G. Waddell, and Francisco Montero.
\newblock Optimization of metabolism: The evolution of metabolic pathways
  towards simplicity through the game of the pentose phosphate cycle.
\newblock {\em J. Theor. Biol.}, 166:201--220, 1994.

\bibitem{Milo:15}
Ron Milo and Rob Phillips.
\newblock {\em Cell Biology by the Numbers}.
\newblock Garland Science, New York, NY, 2015.

\bibitem{Nielsen:2015}
Jens Nielsen.
\newblock Yeast cell factories on the horizon.
\newblock {\em Science}, 349:1050--1051, 2015.

\bibitem{Noor:2010}
Elad Noor, Eran Eden, Ron Milo, and Uri Alon.
\newblock Central carbon metabolism as a minimal biochemical walk between
  precursors for biomass and energy.
\newblock {\em Mol. Cell}, 39:809--820, 2010.

\bibitem{Orth:10}
Jeffrey~D Orth, Ines Thiele, and Bernhard~{\O} Palsson.
\newblock What is flux balance analysis?
\newblock {\em Nat. Biotech.}, 28:245--248, 2010.

\bibitem{papin2004comparison}
Jason~A Papin, Joerg Stelling, Nathan~D Price, Steffen Klamt, Stefan Schuster,
  and Bernhard~O Palsson.
\newblock Comparison of network-based pathway analysis methods.
\newblock {\em Trends Biotechnol.}, 22(8):400--405, 2004.

\bibitem{Planes:20092244}
F.~J. Planes and J.~E. Beasley.
\newblock Path finding approaches and metabolic pathways.
\newblock {\em Discrete Appl. Math.}, 157(10):2244 -- 2256, 2009.
\newblock Networks in Computational Biology.

\bibitem{Planes:01092008}
Francisco~J. Planes and John~E. Beasley.
\newblock A critical examination of stoichiometric and path-finding approaches
  to metabolic pathways.
\newblock {\em Briefings Bioinf.}, 9(5):422--436, 2008.

\bibitem{plasson:2011}
Raphaël Plasson, Axel Brandenburg, Ludovic Jullien, and Hugues Bersini.
\newblock Autocatalyses.
\newblock {\em J. Phys. Chem. A}, 115(28):8073--8085, 2011.

\bibitem{Ricardo:2006}
Alonso Ricardo, Fabianne Frye, Matthew~A. Carrigan, Jeremiah~D. Tipton,
  David~H. Powell, and Steven~A. Benner.
\newblock 2-{H}ydroxymethylboronate as a {R}eagent {T}o {D}etect
  {C}arbohydrates: {A}pplication to the {A}nalysis of the {F}ormose {R}eaction.
\newblock {\em J. Org. Chem.}, 71:9503--9505, 2006.

\bibitem{Ricca:2011}
Emanuele Ricca, Birgit Brucher, and Joerg~H Schrittwieser.
\newblock Multi-enzymatic cascade reactions: Overview and perspectives.
\newblock {\em Adv Synth Catal 353, 2239 – 2262}, 353:2239--2262, 2011.

\bibitem{Ro:2006}
Dae-Kyun Ro, Eric~M. Paradise, Mario Ouellet, Karl~J. Fisher, Karyn~L. Newman,
  John~M. Ndungu, Kimberly~A. Ho, Rachel~A. Eachus, Timothy~S. Ham, James
  Kirby, Michelle C.~Y. Chang, Sydnor~T. Withers, Yoichiro Shiba, Richmond
  Sarpong, and Jay~D. Keasling.
\newblock Production of the antimalarial drug precursor artemisinic acid in
  engineered yeast.
\newblock {\em Nature}, 440:940--943, 2006.

\bibitem{Rohl2017}
Annika R{\"o}hl and Alexander Bockmayr.
\newblock A mixed-integer linear programming approach to the reduction of
  genome-scale metabolic networks.
\newblock {\em BMC Bioinf.}, 18(1):2, Jan 2017.

\bibitem{Ruiz-Mirazo:2014}
Kepa Ruiz-Mirazo, Carlos Briones, and Andr{\'e}s de~la Escosura.
\newblock Prebiotic systems chemistry: New perspectives for the origins of
  life.
\newblock {\em Chem. Rev.}, 114:285--366, 2014.

\bibitem{Savinell:92}
Joanne~M. Savinell and Bernhard~{\O} Palsson.
\newblock Network analysis of intermediary metabolism using linear
  optimization. {I}. {D}evelopment of mathematical formalism.
\newblock {\em J. Theor. Biol.}, 154:421--454, 1992.

\bibitem{schilling2000theory}
Christophe~H Schilling, David Letscher, and Bernhard~{\O} Palsson.
\newblock Theory for the systemic definition of metabolic pathways and their
  use in interpreting metabolic function from a pathway-oriented perspective.
\newblock {\em J. Theor. Biol.}, 203(3):229--248, 2000.

\bibitem{Schuster:94}
S~Schuster and C~Hilgetag.
\newblock On elementary flux modes in biochemical reaction systems at steady
  state.
\newblock {\em J. Biol. Syst.}, 2(02):165--182, 1994.

\bibitem{Stincone:2014}
Anna Stincone, Alessandro Prigione, Thorsten Cramer, Mirjam M~C Wamelink, Kate
  Campbell, Eric Cheung, Viridiana Olin-Sandoval, Nana-Maria Gr{\"u}ning, Antje
  Kr{\"u}ger, Mohammad~Tauqeer Alam, Markus~A Keller, Michael Breitenbach,
  Kevin~M Brindle, Joshua~D Rabinowitz, and Markus Ralser.
\newblock The return of metabolism: biochemistry and physiology of the pentose
  phosphate pathway.
\newblock {\em Biol. Rev.}, 90:927--963, 2015.

\bibitem{lpEnum2}
Garret Swart.
\newblock Finding the convex hull facet by facet.
\newblock {\em Journal of Algorithms}, 6(1):17--48, 1985.

\bibitem{Thodey:2014}
Kate Thodey, Stephanie Galanie, and Christina~D Smolke.
\newblock A microbial biomanufacturing platform for natural and semisynthetic
  opioids.
\newblock {\em Nature Chem Biol}, 10:837--844, 2014.

\bibitem{Zeigarnik:00a}
Andrew~V Zeigarnik.
\newblock On hypercycles and hypercircuits in hypergraphs.
\newblock {\em Discrete Mathematical Chemistry}, 51:377--383, 2000.

\end{thebibliography}
